\documentclass[aps,prc,reprint,groupedaddress]{revtex4-2}
\usepackage{graphicx}
\usepackage{color}

\begin{document}


\title{Measurement of the differential cross sections of $\Sigma^{-}p$ elastic scattering in momentum range of 470 to 850 MeV/$c$}



\author{K. Miwa$^1$, J.K. Ahn$^2$, Y. Akazawa$^3$,  T. Aramaki$^1$, S. Ashikaga$^4$, S. Callier$^{5}$, N. Chiga$^{1}$, S. W. Choi$^{2}$, H. Ekawa$^{6}$, \\
P. Evtoukhovitch$^{7}$, N. Fujioka$^{1}$, M. Fujita$^{8}$, T. Gogami$^{4}$, T. Harada$^{4}$, S. Hasegawa$^{8}$, S. H. Hayakawa$^{1}$, R. Honda$^{3}$, \\
S. Hoshino$^{9}$, K. Hosomi$^{8}$, M. Ichikawa$^{4, 14}$, Y. Ichikawa$^{8}$, M. Ieiri$^{3}$, M. Ikeda$^{1}$, K. Imai$^{8}$, Y. Ishikawa$^{1}$, S. Ishimoto$^{3}$, \\
W. S. Jung$^{2}$, S. Kajikawa$^{1}$, H. Kanauchi$^{1}$, H. Kanda$^{10}$, T. Kitaoka$^{1}$, B. M. Kang$^{2}$, H. Kawai$^{11}$, S. H. Kim$^{2}$, \\
K. Kobayashi$^{9}$, T. Koike$^{1}$, K. Matsuda$^{1}$, Y. Matsumoto$^{1}$, S. Nagao$^{1}$, R. Nagatomi$^{9}$, Y. Nakada$^{9}$, M. Nakagawa$^{6}$, \\
I. Nakamura$^{3}$, T. Nanamura$^{4, 8}$, M. Naruki$^{4}$, S. Ozawa$^{1}$, L. Raux$^{5}$, T. G. Rogers$^{1}$, A. Sakaguchi$^{9}$, T. Sakao$^{1}$, \\
H. Sako$^{8}$, S. Sato$^{8}$, T. Shiozaki$^{1}$, K. Shirotori$^{10}$, K. N. Suzuki$^{4}$, S. Suzuki$^{3}$, M. Tabata$^{11}$, C. d. L. Taille$^{5}$, \\
H. Takahashi$^{3}$, T. Takahashi$^{3}$, T. N. Takahashi$^{10}$, H. Tamura$^{1,  8}$, M. Tanaka$^{3}$, K. Tanida$^{8}$, Z. Tsamalaidze$^{7, 12}$, \\
M. Ukai$^{3}$, H. Umetsu$^{1}$, S. Wada$^{1}$, T. O. Yamamoto$^{8}$, J. Yoshida$^{1}$, K. Yoshimura$^{13}$}

\affiliation{$^1$ Department of Physics, Tohoku University, Sendai 980-8578, Japan.}
\affiliation{$^2$ Department of Physics, Korea University, Seoul 02841, Korea}
\affiliation{$^3$ Institute of Particle and Nuclear Studies (IPNS), High Energy Accelerator Research Organization (KEK), Tsukuba 305-0801, Japan}
\affiliation{$^4$ Department of Physics, Kyoto University, Kyoto 606-8502, Japan}
\affiliation{$^{5}$ OMEGA Ecole Polytechnique-CNRS/IN2P3, 3 rue Michel-Ange, 75794 Paris 16, France}
\affiliation{$^{6}$ High Energy Nuclear Physics Laboratory, RIKEN, Wako, 351-0198, Japan}
\affiliation{$^{7}$ Joint Institute for Nuclear Research (JINR), Dubna, Moscow Region 141980, Russia}
\affiliation{$^8$ Advanced Science Research Center (ASRC), Japan Atomic Energy Agency (JAEA), Tokai, Ibaraki 319-1195, Japan}
\affiliation{$^9$ Department of Physics, Osaka University, Toyonaka 560-0043, Japan}
\affiliation{$^{10}$ Research Center for Nuclear Physics (RCNP), Osaka University, Ibaraki 567-0047, Japan}
\affiliation{$^{11}$ Department of Physics, Chiba University, Chiba 263-8522, Japan}
\affiliation{$^{12}$ Georgian Technical University (GTU), Tbilisi, Georgia}
\affiliation{$^{13}$ Department of Physics, Okayama University, Okayama 700-8530, Japan}
\affiliation{$^{14}$ Meson Science Laboratory, Cluster for Pioneering Research, RIKEN, Wako, 351-0198, Japan}

\collaboration{J-PARC E40 Collaboration}

\date{\today}

\begin{abstract}
A high statistics $\Sigma p$ scattering experiment is performed at the J-PARC Hadron Experimental Facility.
Momentum-tagged $\Sigma^{-}$s running in a liquid hydrogen target are accumulated by detecting the $\pi^{-}p \to K^{+}\Sigma^{-}$ reaction  with a high intensity $\pi^{-}$ beam of 20 M/spill.
The differential cross sections of the $\Sigma^{-}p$ elastic scattering were derived with a drastically improved accuracy by identifying approximately 4,500 events from 1.72 $\times$ $10^{7}$ $\Sigma^{-}$.
The derived differential cross section shows a clear forward-peaking angular distribution for a $\Sigma^{-}$ momentum range from 470 to 850 MeV/$c$.
The accurate data will impose a strong constraint on the theoretical models of the baryon-baryon interactions.
\end{abstract}


\maketitle



\section{Introduction}

The interactions between octet baryons, namely, baryon-baryon ($BB$) interactions including hyperon-nucleon ($YN$) and hyperon-hyperon ($YY$) interactions have been intensively studied both theoretically and experimentally. 
The Nijmegen \cite{Rijken:1999, Nagels:2019} and J\"{u}lich \cite{Haidenbauer:2005} groups constructed their own $BB$ interaction models with a unified description based on the boson-exchange picture assuming the flavor SU(3) symmetry.
The role of quarks in the $BB$ interactions was studied using the Quark Cluster Model (QCM) by considering the effect of the Pauli principle for quarks and the color magnetic interaction \cite{Oka:1986}.
QCM  predicts characteristic features of the short-range region such as a strongly repulsive core or an attractive interaction depending on the spin and flavor configuration of the quarks in the system.
A realistic description including the quark degree of freedom  by the Kyoto-Niigata group \cite{Fujiwara:2007} incorporates an effective meson exchange potential in the QCM to represent the middle and long range interactions.
These predictions are reproduced by lattice QCD simulations, which become a powerful theoretical tool to derive the $YN$ and $YY$ potentials from the first principle in QCD \cite{Aoki:2012, Inoue:2012, Nemura:2018}.
A modern description based on the chiral effective field theory ($\chi$EFT) was extended to the $YN$ sector up to the next leading order \cite{Haidenbauer:2013, Haidenbauer:2020}.
To test and improve these  theoretical models of the $BB$ interactions, important experimental inputs must be the two-body scattering data between a hyperon and a proton.
In the case of the $NN$ interaction (nuclear force), the $pp$ and $np$ scattering data played essential roles to establish realistic models of nuclear force \cite{Machleidt:2001, Stoks:1994, Wiringa:1995}.
However, no experimental progress in hyperon-proton scattering has been made since the 1970s because of experimental difficulties stemming from the low intensity of the hyperon and its short lifetime \cite{Sechi-zorn:1968, Alexander:1968, Kadyk:1971, Hauptman:1977, Engelmann:1966, Eisele:1971, Stephen:1970, Kondo:2000, Kanda:2005}.
Therefore, historically, $BB$ interactions have been examined from hypernuclear data because their binding energies and energy levels reflect the $YN$ interactions \cite{Hashimoto:2006, Yamamoto:2010}.
Since the observation of a massive neutron star with a two-solar mass \cite{Demorest:2010}, the existence of a three-body repulsive interaction including hyperons has been discussed as a possible source of supporting such massive stars \cite{Yamamoto:2014}.
To derive the properties of the $YNN$ three-body interaction from the hypernuclear structure, the $YN$ two-body interaction should be determined from the two-body system to eliminate uncertainties from the many-body effects in the hypernuclear system.
For these reasons,  the construction of a realistic two-body $YN$ interaction via high statistics hyperon-proton  scattering data is crucially important.

As a first step, 
a new $\Sigma p$ scattering experiment was performed in J-PARC to provide accurate differential cross sections of the $\Sigma^{+}p$, $\Sigma^{-}p$ elastic scatterings and $\Sigma^{-}p \to \Lambda n$ inelastic scattering.
Theoretically, the $\Sigma N$ interaction is predicted as strongly spin-isospin dependent.
The only observed  $\Sigma$ hypernucleus ($^{4}_{\Sigma}$He)  \cite{Nagae:1998} is bound by the attractive interaction in the ($I=1/2$, $S=1$) channel.
However,  the spin and isospin averaged $\Sigma$-nucleus potential was confirmed as strongly repulsive from $\Sigma$ quasi-free production spectra in several nuclei \cite{Saha:2004, Harada:2018}.
To examine the $\Sigma N$ interaction for each spin-isospin channel more quantitatively, the systematic measurement of these three scattering channels is important.
In this paper, we present the differential cross section of the $\Sigma^{-}p$ elastic scattering in the $\Sigma^{-}$ momentum region ranging from 470 to 850 MeV/$c$ as the first result of the systematic measurement.
The existing differential cross sections of  the $\Sigma^{-} p$ scattering are limited to the $S$-wave dominant region of a beam momentum around 300 MeV/$c$ \cite{Engelmann:1966, Eisele:1971, Stephen:1970} except for a few higher momentum data with a large uncertainty \cite{Kondo:2000}.
Therefore, there are no data to determine the $P$- and higher partial waves of the $\Sigma p$ channel \cite{Haidenbauer:2013, Haidenbauer:2020}.
Particularly, in the $\Sigma^{-}p$ channel, all theoretical models predict a large angular dependence in the differential cross section in the $\Sigma^{-}$ momentum range higher than 500 MeV/$c$ owing to the higher partial wave contribution.
The precise measurement of the differential cross section of the $\Sigma^{-}p$ elastic scattering in this momentum region is indispensable  to impose a strong constraint on the theoretical models.

The present article is organized as follows.
In Sec. \ref{sec_exp}, an outline of the experiment is described.
In Sec. \ref{sec_analysis1}, an analysis to identify the $\Sigma^{-}p$ scattering is described.
At first, the $\Sigma^{-}$ particles are identified from a missing mass spectrum of the $\pi^{-}p \to K^{+}X$ reaction.
Then, an analysis of so called CATCH system, which is a main detector to identify the $\Sigma^{-}p$ scattering, is described.
In this analysis, the $\Sigma^{-}p$ elastic scattering events are successfully identified.
In Sec. \ref{sec_ana2}, an analysis to derive  differential cross sections of the $\Sigma^{-}p$ elastic scattering is described in detail.
Finally, the differential cross section is reported and is compared with theoretical calculations.
A summary follows in Sec. \ref{sec_summary}.

\section{Experiment}\label{sec_exp}

\begin{figure}[]
\centerline{\includegraphics[width=0.5\textwidth]{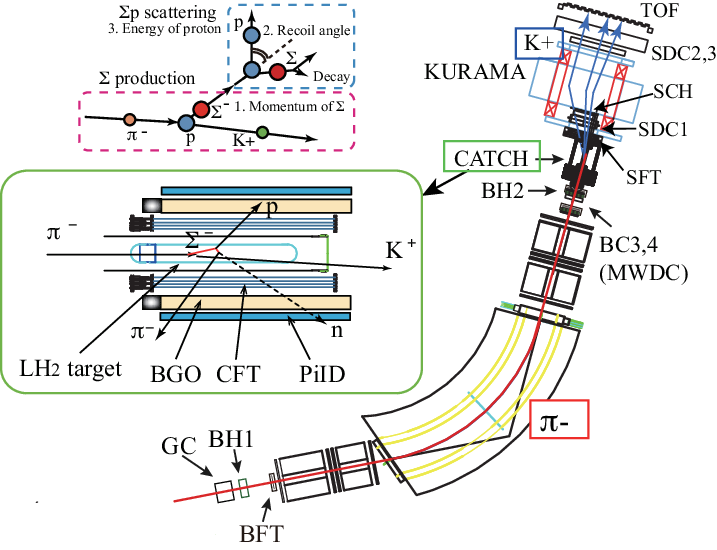}}
  \caption{Experimental concept of the $\Sigma p$ scattering experiment and experimental setup with an enlarged figure around the LH$_{2}$ target.
Two successive two-body reactions of  the $\Sigma^{-}$ production ($\pi^{-} p \to K^{+} \Sigma^{-}$) and the $\Sigma^{-} p$ scattering ($\Sigma^{-} p \to \Sigma^{-} p$) are detected.
The momentum of the produced $\Sigma$ particles is obtained from the
momenta of $\pi$ beam particles and scattered $K^{+}$ measured by the K1.8 beam line spectrometer and
KURAMA spectrometer, respectively.
The beam line spectrometer consists of two hodoscopes (BH1 and BH2) and three position detectors (BFT, BC3 and BC4). 
In the KURAMA spectrometer, five position detectors (SFT, SDC1, SCH, SDC2 and SDC3) and a TOF counter are used. 
The $\Sigma p$ scattering events are detected by the CATCH system, which surrounds the LH$_{2}$ target.
  }
  \label{K18beamline_spectrometer_KURAMA_w_CATCH}
\end{figure}

The $\Sigma p$ scattering experiment (J-PARC E40) \cite{Miwa:2019, Miwa:2020} was performed at the K1.8 beam line in the J-PARC Hadron Experimental Facility.
A 1.33 GeV/$c$ $\pi^{-}$ beam of $2.0 \times10^{7}$/spill was produced from a 30 GeV proton beam with a spill cycle of 5.2 s and a beam duration of 2 s.
Fig. \ref{K18beamline_spectrometer_KURAMA_w_CATCH} shows the experimental concept and setup with an enlarged view around a liquid hydrogen (LH$_{2}$) target.
$\Sigma^{-}$ particles are produced by the $\pi^{-}p \to K^{+}\Sigma^{-}$ reaction and the produced $\Sigma^{-}$ running in the LH$_{2}$ target interacts.
The momentum of each $\Sigma^{-}$ can be tagged as the missing momentum calculated from the momenta of the $\pi^{-}$ beam and outgoing $K^{+}$ analyzed by the K1.8 beam line spectrometer \cite{Takahashi:2012} and  forward magnetic  spectrometer (KURAMA), respectively.
A recoil proton knocked out by the $\Sigma^{-} p$ scattering is detected by the CATCH system, which consisted of a cylindrical scintillation fiber tracker (CFT), a BGO calorimeter (BGO) and a scintillator hodoscope (PiID) coaxially arranged from the center  outwards \cite{Akazawa:2019}.
The $\Sigma^{-}p$ elastic scattering can be identified by checking the kinematical consistency between the scattering angle and kinetic energy of the recoil proton measured by CFT and BGO, respectively.
The spectrometers  and  CATCH are described in detail in the following paragraphs. 

A high intensity $\pi^{-}$ beam of approximately 2$\times 10^{7}$ /spill was used to accumulate a large number of $\Sigma^{-}$ particles. 
The beam momentum was reconstructed event by event with the K1.8 beam-line spectrometer, which consisted of QQDQQ magnets and position detectors (a fiber detector (BFT \cite{Honda:2015}) and drift chambers (BC3,4)) placed upstream and downstream of the spectrometer magnets.
The liquid hydrogen was filled in a target container of a cylinder having a diameter of 40 mm and length of 300 mm with half-sphere end-caps for both edges as shown in Fig. \ref{K18beamline_spectrometer_KURAMA_w_CATCH}.
The target container was made with a Mylar sheet of 0.25 mm thickness.
A vacuum window around the target region was made by a CFRP cylinder with a diameter of 80 mm  and thickness of 1 mm to minimize the amount of material.
Outgoing particles produced at the LH$_{2}$ target by the $\pi^{-} p$ reaction were detected by the KURAMA spectrometer placed downstream of the LH$_{2}$ target.
The charged particles were bent in a magnetic field of 0.78 T in the KURAMA magnet and its momentum was analyzed by measuring the positions  upstream and downstream of the magnet
with position detectors (fiber tracker (SFT) and three drift chambers (SDC1, 2 and 3)).
The time of flight from the the BH2 counter placed at the upstream of the LH$_{2}$ target was measured by a hodoscope (TOF) placed downstream of the SDC3.
The spectrometer acceptance for $K^{+}$ for the $\pi^{-}p \to \Sigma^{-} K^{+}$ reaction was approximately 4\%.
The flight length from the LH$_{2}$ target to TOF  was approximately 3 m and the typical survival rate of $K^{+}$ was 59\%.
The large acceptance and short flight length were advantages of the KURAMA spectrometer to accumulate a large number of  the $\Sigma^{-}$ particles.
An aerogel cherenkov counter (SAC) and a fine segmented hodoscope (SCH) were installed at the front part of the KURAMA spectrometer for the trigger function.
SAC, whose refractive index is 1.10, was used to reject $\pi^{+}$ at the trigger level.
The hit combination between SFT, SCH, and TOF was used to select the momentum range of 0.6 $\sim$ 1.1 GeV/$c$ roughly, which was typical momentum of $K^{+}$ for the $\Sigma^{-}$ production.

The second component of the detector system is CATCH, which is used to identify the $\Sigma^{-} p$ scattering that occurs between the running $\Sigma^{-}$ and protons in the LH$_{2}$ target.
CATCH consists of a cylindrical fiber tracker (CFT), a BGO calorimeter (BGO) and a scintillator hodoscope (PiID)\cite{Akazawa:2019}.
The CFT consists of eight layers of cylindrical fiber layers whose radii range from 50 mm to 85 mm with 5 mm intervals.
The CFT has two different layer configurations, that is, the $\phi$ and $uv$ layers.
In the $\phi$ layer, scintillation fibers with a diameter of 0.75 mm were placed in parallel with the beam direction around the cylinder position.
In the $uv$ layer, scintillation fibers with the same diameter were placed in a spiral shape along the cylinder position and the $u$ and $v$ layers had tilt angle directions opposite to the beam direction.
The CFT consisted of eight layers of the alternate fiber configuration, namely , $u1, \phi 1, v2, \phi 2, u3, \phi 3, v4$  and $\phi 4$ layers.
Trajectories of the charged particles from the LH$_{2}$ target were reconstructed three-dimensionally. 
BGO was placed around the CFT to measure the kinetic energy of the recoil proton by stopping in the calorimeter.
BGO consisted of 22 BGO crystals whose size was 400  mm ($l$) $\times$ 30  mm ($w$) $\times$ 25 mm ($t$), each of which was read out by connecting a photomultiplier tube at the downstream surface.
Finally, PiID, which consisted of  29 segments of the scintillation counter,  was placed  outside the BGO to check whether the charged particle was stopped in the BGO or not.
The experiment was performed for approximately 20 days in February, 2019.
In addition to the physics data taking, $pp$ scattering data with various proton beam momenta ranging from 450 to 800 MeV/$c$ were also collected for the energy calibrations of the CFT and BGO.
These data were also used to estimate the detection efficiency of CATCH.

\section{Analysis 1 : Identification of the $\Sigma^{-}p$ scattering event} \label{sec_analysis1}

Analysis of the $\Sigma^{-}p$ scattering consists of three components.
First,  the momentum-tagged $\Sigma^{-}$ particles are identified from the analysis of the beam-line and KURAMA spectrometers.
Subsequently,  the $\Sigma^{-}p$ scattering events are identified from the analysis of CATCH for the $\Sigma^{-}$ production events.
Finally,  the differential cross section is derived.
In the following sub sections, we will explain the first two components.
The analysis of the derivation of the differential cross section is described in the next section.

\subsection{Analysis of the momentum-tagged $\Sigma^{-}$ particles}

$\Sigma^{-}$ particles are identified by the missing mass spectrum of the $\pi^{-}p \to K^{+}X$ reaction.
The particle identification and the momentum reconstruction for both the $\pi^{-}$ beam and outgoing $K^{+}$ should be performed.

\begin{figure}[]
  \centerline{\includegraphics[width=0.5\textwidth]{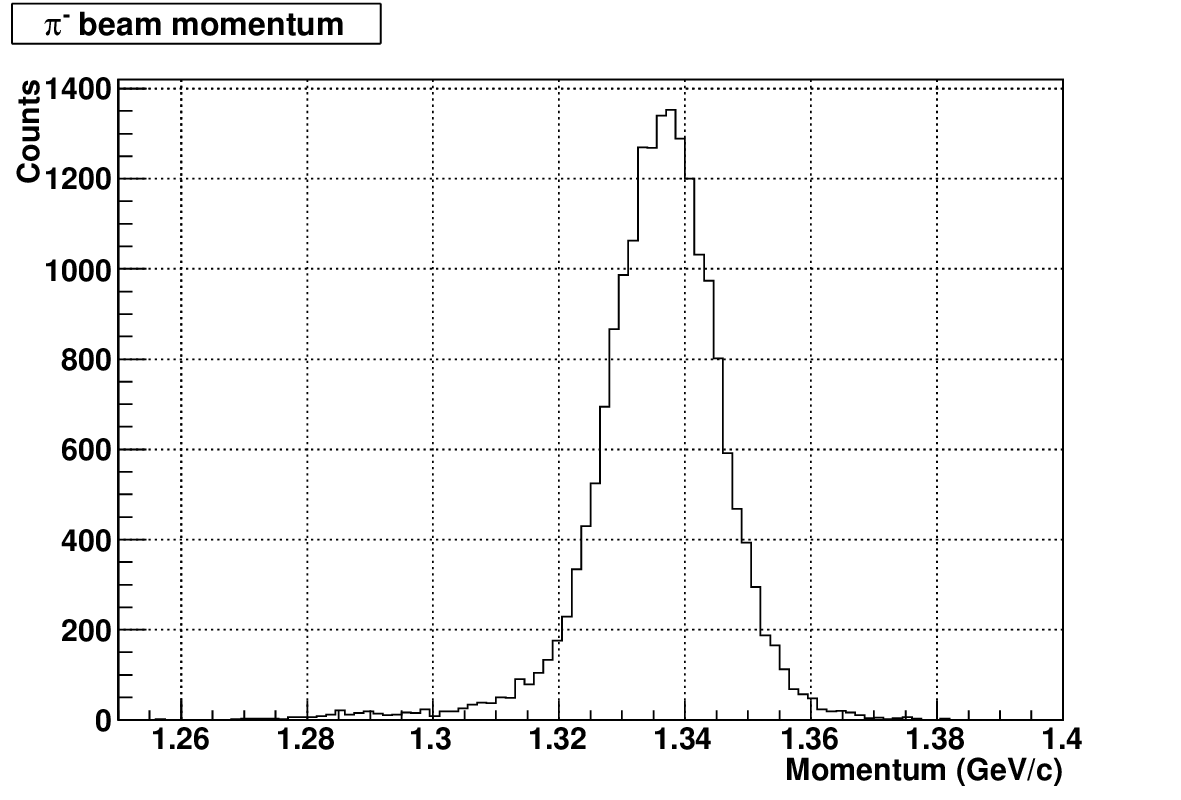}}
  \caption{ Reconstructed momentum for the $\pi^{-}$ beam. 
  }
  \label{showK18Mom}
\end{figure}

The momenta of the beam particles were analyzed by the K1.8 beam-line spectrometer event by event.
Straight tracks at the downstream of the spectrometer magnet were reconstructed using BC3, 4.
The beam momentum was reconstructed by connecting the downstream track to the hit position at the BFT located at the upstream of the spectrometer magnet with a third-order transfer matrix.
Fig. \ref{showK18Mom} shows the reconstructed momentum for the $\pi^{-}$ beam.

The outgoing particles produced at the LH$_{2}$ target by the $\pi^{-}p$ reaction were analyzed by the KURAMA spectrometer.
The straight tracks were obtained for both the upstream and  downstream of the KURAMA magnet and these tracks were connected with a Runge-Kutta method \cite{Myrheim:1979} considering the equation of motion under the magnetic field interpolated from the field map of the KURAMA magnet.
The momentum resolution was expected to be $\sigma/P  \sim 10^{-2}$.

\begin{figure}[]
  \centerline{\includegraphics[width=0.5\textwidth]{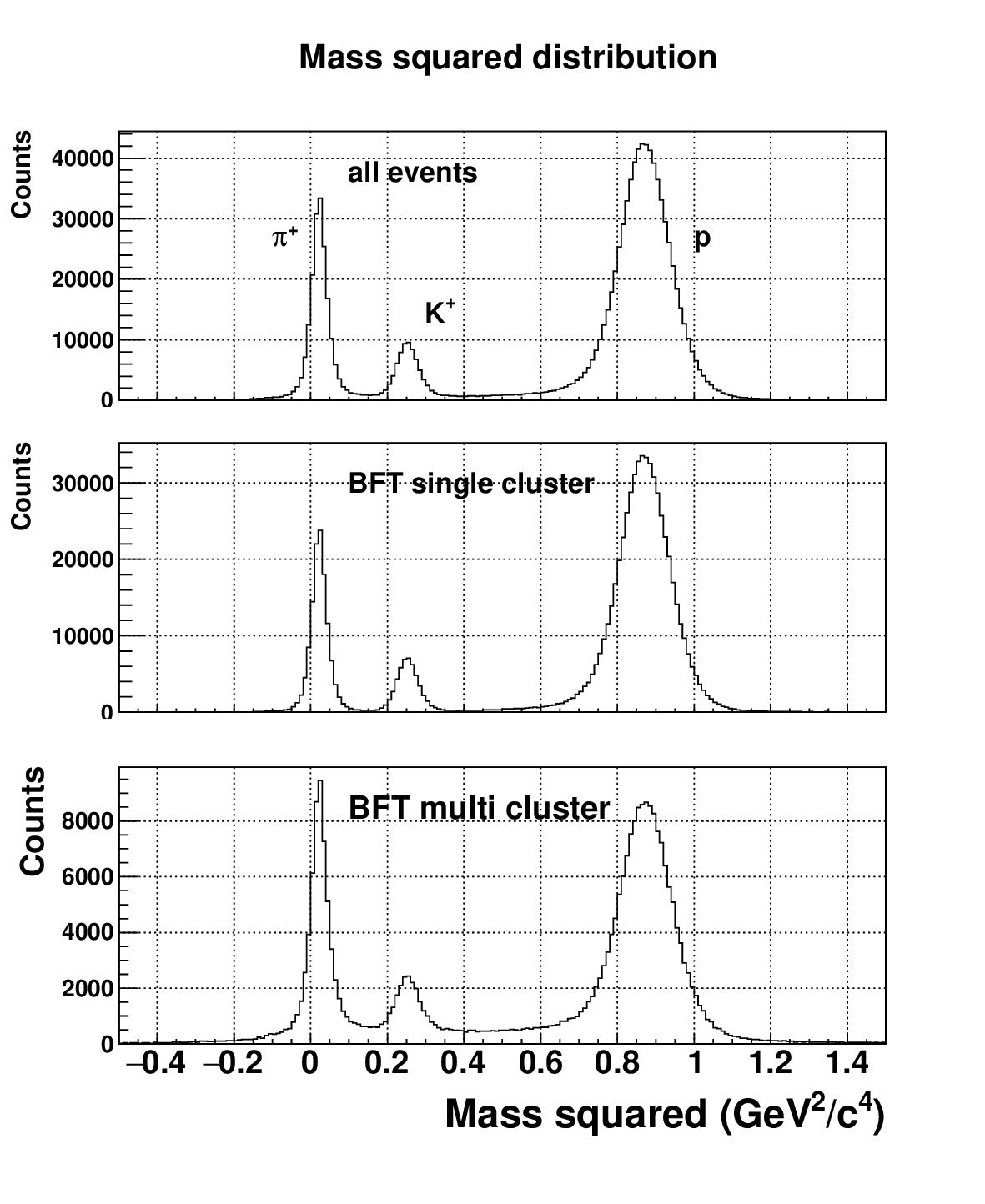}}
  \caption{ Reconstructed mass squared distribution for all events (top),  BFT single cluster events (middle) and BFT multi cluster event (bottom).
  }
  \label{showMassWithBftCluster}
\end{figure}

 \begin{figure*}[]
  \centerline{\includegraphics[width=0.8\textwidth]{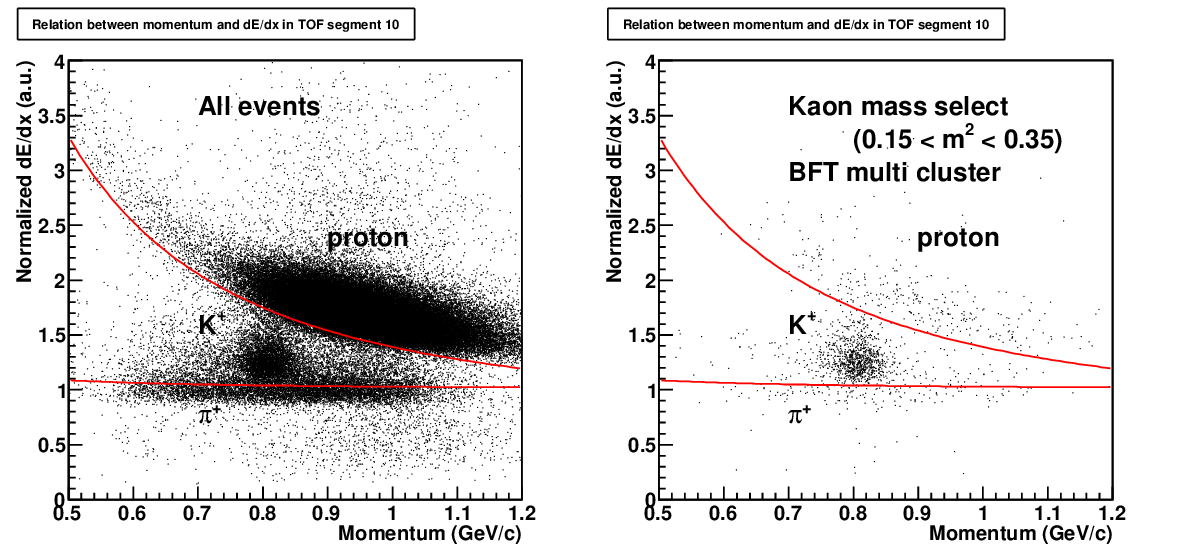}}
  \caption{ Correlation between the momentum and  energy deposit per unit length at the TOF, which was normalized as one for $\pi^{+}$.
Left and right figures show the correlations of all events and  the $K^{+}$  mass region for BFT multi-cluster event, respectively.
Region between two red lines was selected as the $K^{+}$ region for the BFT-multi cluster event.
  }
  \label{showTofdEAndMomForEps.eps}
\end{figure*}

Fig. \ref{showMassWithBftCluster} shows the mass squared ($m^{2}$) distribution reconstructed by the KURAMA spectrometer for different beam-multiplicity conditions.
By measuring the hit cluster number at BFT where the beam envelope became wide, the beam multiplicity within a $\pm$5 ns time gate could be estimated.
The top figure in Fig. \ref{showMassWithBftCluster} shows $m^{2}$ for all events.
There were miscalculated events in the $m^{2}$ spectrum as a constant background between the $\pi^{+}$ and  proton peaks.
This background also occuered under the $K^{+}$ peak.
These background events were attributed to the misidentification of the hit timing at BH2, which determined a start timing of the time-of-flight measurement, owing to the high beam multiplicity.
 As shown in the middle figure in Fig. \ref{showMassWithBftCluster}, the contamination of the miscalculated events was quite suppressed for the BFT single-cluster event,
 whereas it was enhanced for the BFT multi-cluster events as shown in the bottom figure in Fig.  \ref{showMassWithBftCluster}.
 Because the beam was focused at the LH$_{2}$ target, multiple beam particles could hit the same BH2 segment within a short time interval.
This caused the misidentification of the timing at BH2, which resulted in the miscalculation of $m^{2}$.
 To suppress the miscalculated events, the energy deposit information at  TOF was used.
The figure on the left in Fig. \ref{showTofdEAndMomForEps.eps} shows the correlation between the momentum and energy deposit per unit length, which was normalized as one for $\pi^{+}$.
The loci corresponding to $\pi^{+}$, $K^{+}$ and the proton could be identified.
We set the separation lines between $\pi^{+}$ and $K^{+}$ and between $K^{+}$ and the proton as shown by the red lines in Fig. \ref{showTofdEAndMomForEps.eps}.
This additional cut was applied only to the BFT multiple-cluster events as shown in the figure on the right in Fig. \ref{showTofdEAndMomForEps.eps}, because this cut also rejected some of the $K^{+}$ events.
Finally, $K^{+}$ was selected from the $m^{2}$ gate, which was optimized depending on the $K^{+}$ momentum, and from the $dE/dx$ gate at the TOF for the BFT multiple-cluster events.
The contamination rate of the miscalculated events was estimated to be 7.5\% for the selected $K^{+}$ events.

 \begin{figure*}[]
  \centerline{\includegraphics[width=0.8\textwidth]{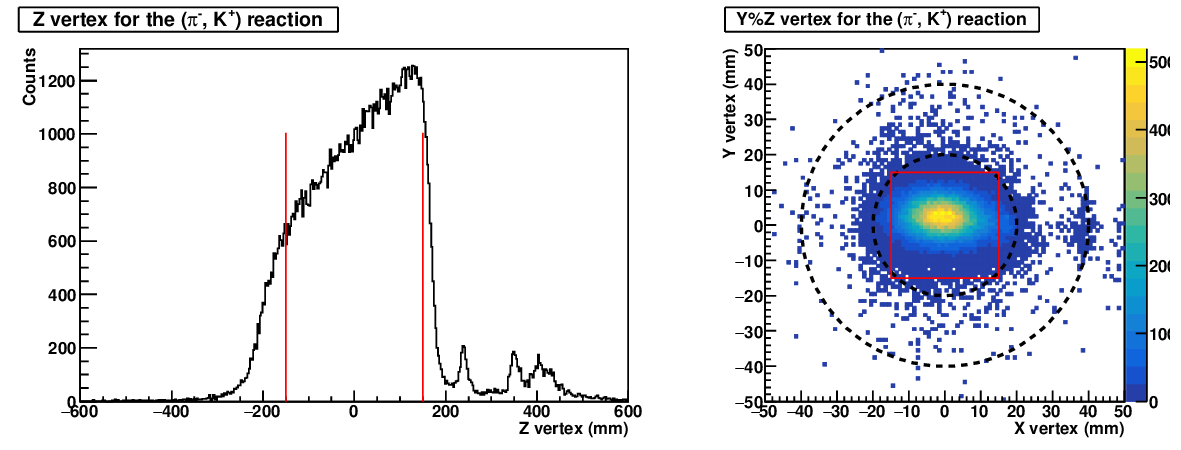}}
  \caption{ Vertex distribution between $\pi^{-}$ beam and outgoing $K^{+}$.
  The left figure shows the $z$ vertex distribution. 
  The $z$ vertex region from $-$150 mm to 150 mm, which is shown by the red lines, was selected for the $\Sigma^{-}$ production event at the LH$_{2}$ target.
  The right figure shows the scatter plot between the $x$ and $y$ vertices. The dotted lines show the envelopes of the target container and vacuum window.
  The red-box region was selected to suppress the contamination of the reaction at the target container.
  }
  \label{showVertexSigmaMinus}
\end{figure*}

\begin{figure}[]
\begin{center}
  \centerline{\includegraphics[width=0.45\textwidth]{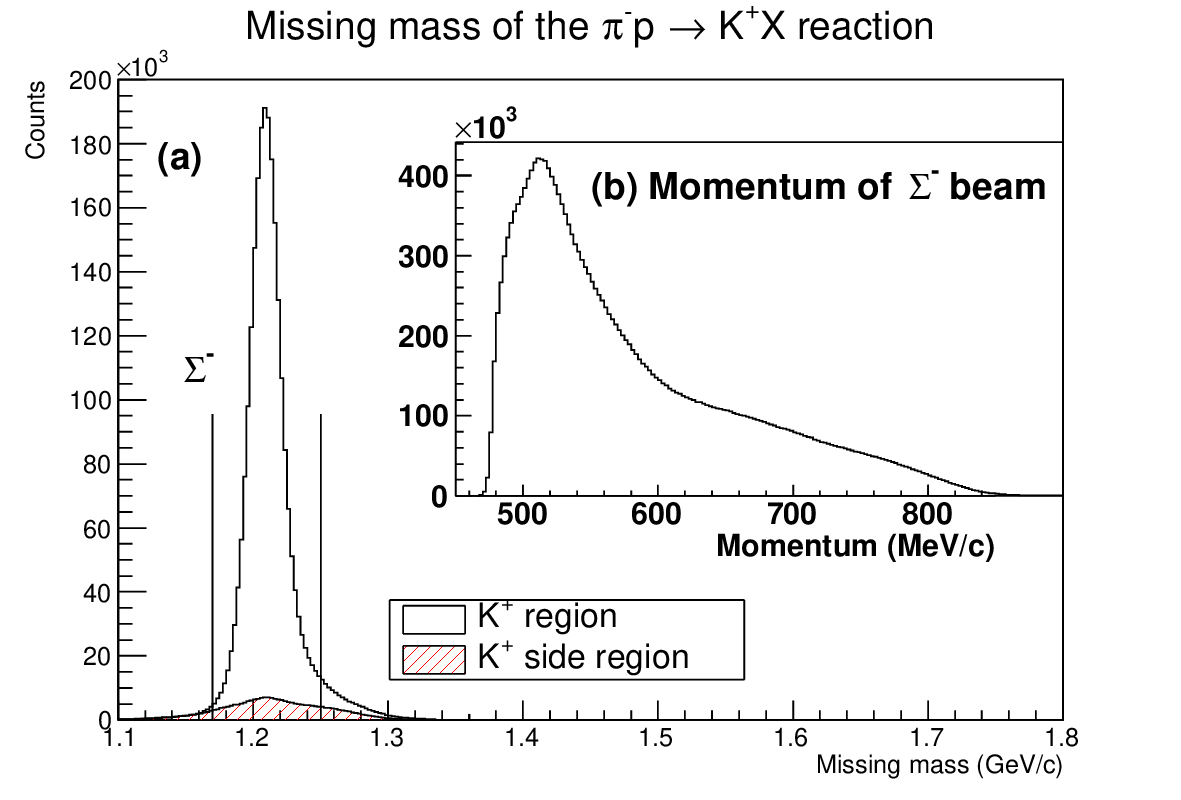}}
   \caption{ (a)  Missing mass spectra of the $\pi^{-}p \to K^{+} X$ reaction for $K^{+}$ events (open histogram) and side-band events of $K^{+}$ (filled histogram) to estimate the effect of the contamination of the miscalculated event under the $K^{+}$ region in mass square spectrum.
The two lines show the selected area for the $\Sigma^{-}$ particles.
(b)  $\Sigma^{-}$ momentum reconstructed as the missing momentum of the $\pi^{-}p \to K^{+} \Sigma^{-}$ reaction.
   }
      \label{showSigmaWithSB}
   \end{center}
\end{figure}

$\Sigma^{-}$ particles were identified from the missing mass spectrum of the $\pi^{-}p \to K^{+}X$ reaction using the reconstructed momenta for the $\pi^{-}$ beam and outgoing $K^{+}$.
Fig. \ref{showVertexSigmaMinus} shows the vertex distributions between the $\pi^{-}$ beam and outgoing $K^{+}$.
The figure on the left in Fig. \ref{showVertexSigmaMinus} shows the $z$ vertex distribution.
The vertex image for the LH$_{2}$ target can be identified from $-$200 mm to 150 mm, whereas the peaks around $z$=240 mm and 400 mm corresponded to the interaction at the vacuum window and SFT, respectively.
For the $\Sigma^{-}p$ scattering analysis, the $z$ vertex region from $-150$ mm to 150 mm as shown by the two lines was used  to satisfy the acceptance for CATCH.
The figure on the right in Fig. \ref{showVertexSigmaMinus} shows the scatter plot between the $x$ and $y$ vertices where the envelopes of the target container and vacuum window are overlaid by the dotted circles.
The beam size for the $x$ direction was slightly wider than the target size, whereas the beam was well focused for the $y$ direction.
The contribution of the vacuum window and target container could be identified.
To suppress such contamination, the $x$ and $y$ vertex regions were selected  from $-15$ mm to $15$ mm as shown in the red box. 
The vertex resolution by the spectrometers was estimated by comparing the vertex from the spectrometer analysis with that obtained from the CATCH analysis for the multi-particle events in CATCH from the same reaction vertex.
The $z$ vertex resolution depended on the scattering angle ($\theta$) and the typical resolutions were $\sigma_{z}$=20 mm and $\sigma_{z}$=9 mm for $\theta=10^{\rm o}$ and  $\theta=20^{\rm o}$, respectively.
The $x$ and $y$ vertex resolutions were estimated as $\sigma_{x}$=3.0 mm and $\sigma_{y}$=3.6 mm, respectively and the angular dependence was negligible.
These vertex resolutions were taken into account in the simulation study to estimate the analysis cut efficiencies.
The missing mass spectrum of the $\pi^{-}p \to K^{+}X$ reaction is shown in Fig. \ref{showSigmaWithSB} (a).
A clear peak corresponding to $\Sigma^{-}$ could be identified.
As we mentioned the $K^{+}$ selection was contaminated due to the miscalculation of the time-of-flight caused by the multiple beam events.
This contribution was studied by selecting the side-band region of $K^{+}$ , that is, $0.1 < m^{2} ({\rm GeV}/c) < 0.15$ and  $0.38 < m^{2} ({\rm GeV}/c) < 0.5$ and these side-band events were scaled to the real background contamination under the $K^{+}$ peak.
The histogram filled by slashed lines shows the missing mass spectrum for the side-band events assuming the outgoing particles are $K^{+}$.
The tail toward the lower mass region less than 1.17 GeV/$c^{2}$ was well reproduced by this miscalculated events.
We selected the $\Sigma^{-}$ particles from the 1.17 to 1.25 GeV/$c^{2}$ region.
When the differential cross section, which is related to the number of $\Sigma^{-}$ particles,  was derived, the contribution of the contamination was subtracted using the side-band events.
In total, the $\Sigma^{-}$ particles of 1.62 $\times 10^{7}$, which is approximately 100 times more than that in a past experiment \cite{Kondo:2000}, were accumulated after subtracting the contamination.
The reconstructed $\Sigma^{-}$ momentum ranged from 470 to 850 MeV/$c$ continuously as shown in Fig. \ref{showSigmaWithSB} (b).

\subsection{Calibration of CATCH  with $pp$ scattering}


Before describing the analysis for the $\Sigma^{-}p$ scattering, we describe the basic analyses of CATCH such as the energy calibration of BGO.
These analyses were performed using the $pp$ scattering data taken by providing proton beams with different momentum conditions from 450 to 850 MeV/$c$.

\begin{figure}[]
  \centerline{\includegraphics[width=0.5\textwidth]{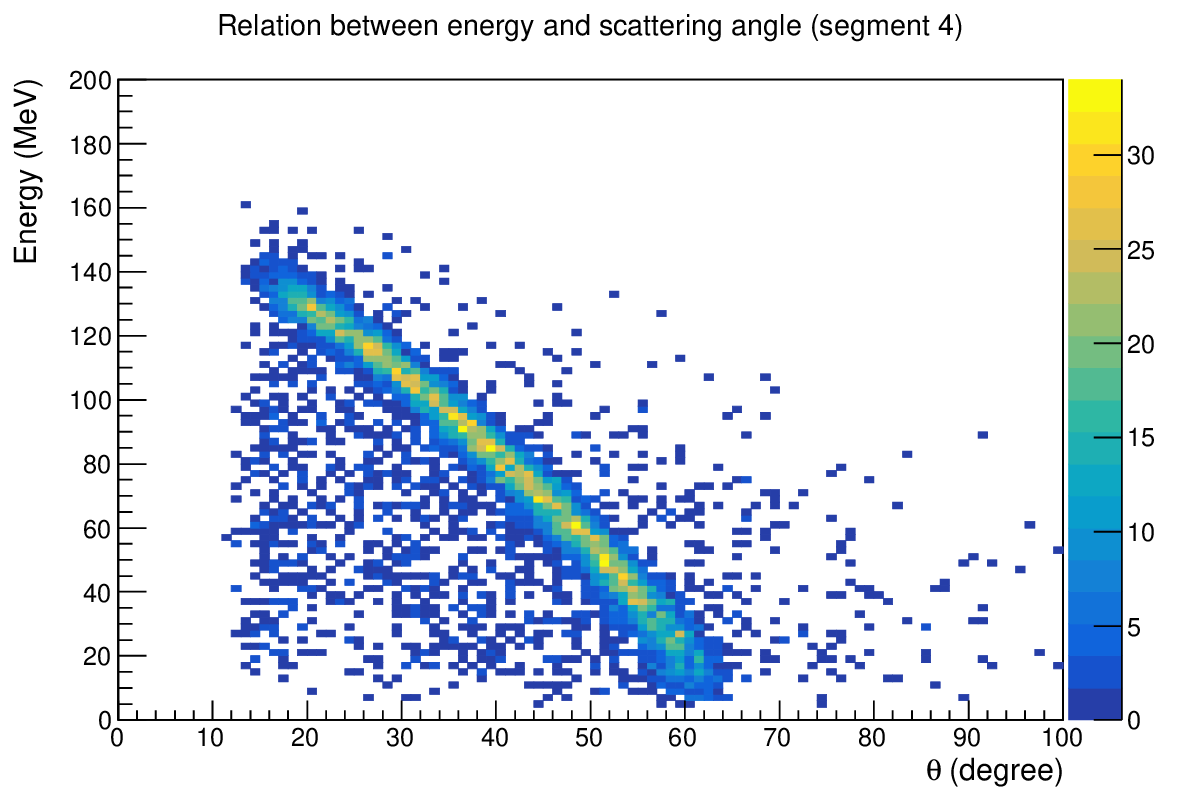}}
  \caption{ Correlation between the scattering angle and kinetic energy measured by the CFT and  BGO for the  $pp$ scattering data with a 600 MeV/$c$ proton beam.
Locus corresponding to the elastic scattering can be identified.
  }
  \label{showBGO_E_Theta}
\end{figure}

The energy calibration of BGO was performed  using the $pp$ elastic scattering where the correlation between the scattering angle and kinetic energy of the proton existed.
The kinetic energy of the proton can be calculated from the scattering angle measured by the CFT tracking.
The correlation between the calculated kinetic energy and pulse height of the BGO was measured for each scattering angle.
The energy calibration of BGO was performed by fitting this correlation.
Fig. \ref{showBGO_E_Theta} shows the correlation between the scattering angle and kinetic energy measured as a sum of the energy deposits in the CFT and BGO for the $pp$ scattering data with the 600 MeV/$c$ proton beam.
The locus corresponding to the $pp$ elastic scattering can be identified.
The energy dependence of the energy resolution of the BGO was estimated from the width of the $pp$ scattering locus for different beam momenta.
The energy resolution is then expressed  by the following equation,
\begin{equation} \label{eq_bgo_reso}
\sigma ({\rm MeV}) = 0.5\sqrt{\frac{E ({\rm MeV})}{80}} + 4.0 ,
\end{equation}
where $E$ represents the kinetic energy of a proton.

The energy calibration of CFT was performed in the same manner by considering the saturation effect of the readout photon sensor (MPPC).
The angular resolution of the CFT was estimated as 1.5 degrees ($\sigma$) from the width of the opening angle of two protons in the $pp$ scattering which was constant at approximately 90$^{\rm o}$ kinematically.
The vertex resolution by the CFT tracking was studied by reconstructing the image of the target container from the vertex of the two-proton tracks in CATCH for the $\pi^{-}$ beam run.
By requiring the two protons in CATCH,  nuclear components such as the target container were enhanced even for the LH$_{2}$ target filled with liquid hydrogen.
The $z$ vertex resolution was estimated as 1.8 mm in $\sigma$ and the $x$ and $y$ vertex resolutions were  estimated as 1.9 mm in $\sigma$ for both directions.
The angular resolution of CFT was also included in the simulation.

\subsection{Analysis of the $\Sigma^{-}p$ scattering} \label{sec_ana_sigmaMp}

\begin{figure}[]
  \centerline{\includegraphics[width=0.5\textwidth]{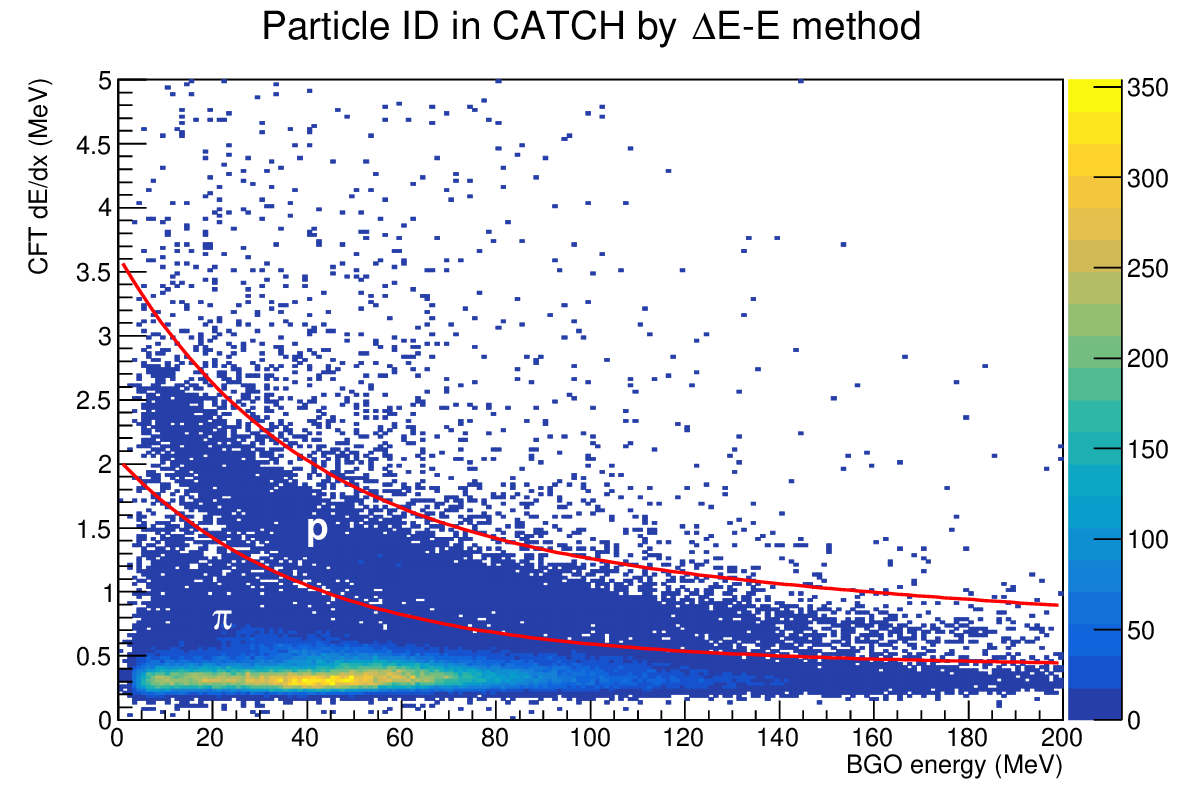}}
  \caption{ Correlation between the total energy measured by the BGO and the energy deposit normalized to the unit length in CFT.
  The two curves show the selection region for the proton.
  Many of the $\pi$s were not stopped in  the BGO and penetrated with only a part of the energy deposit.
  }
  \label{showDeltaE_E_SigmaMinus_CutPos.eps}
\end{figure}

For the analysis of the $\Sigma^{-}p$ scattering, a proton was searched for using CATCH in coincidence with the $\Sigma^{-}$ production.
The particle identification between the $\pi$ and proton was performed by the $dE$-$E$ method between the partial  energy deposit ($dE$) in the CFT and the total energy deposit ($E$) in the BGO as shown in Fig. \ref{showDeltaE_E_SigmaMinus_CutPos.eps}.
The locus corresponding to a proton could be identified and the region defined by the two lines in Fig. \ref{showDeltaE_E_SigmaMinus_CutPos.eps} was selected as the proton event.
For $\pi$, the thickness of BGO was not sufficient to stop the $\pi$s and only the direction was obtained by the CFT tracking.
The momentum estimation of $\pi$ is described later.
For the proton,  its kinetic energy and the direction could be measured by CATCH.
Because $\Sigma^{-}$ has no decay channel to a proton, the detected proton is a signal of the secondary $\Sigma^{-}p$ reactions including the $\Sigma^{-}p$ elastic scattering and $\Sigma^{-}p \to \Lambda n, \Sigma^{0}n$ inelastic scatterings.
However, many of these protons  originated from the secondary $np$ and $\pi^{-}p$ reactions  after the $\Sigma^{-} \to n \pi^{-}$ decay.
The secondary reactions with the proton in the final state in coincidence with the $\Sigma^{-}$ production are summarized in Table \ref{relation_secondary_reaction}.

\begin{table*}
\begin{center}
\caption{List of secondary reactions with a proton in the final state in coincidence with the $\Sigma^{-}$ production.}
\label{relation_secondary_reaction}
\begin{tabular}{cc}
\\
\hline\hline
Reaction &   \\
\hline\hline 
$\Sigma^{-}p$ reaction &   \\
\hline\hline 
$\Sigma^{-}p$ elastic scattering & $\Sigma^{-}p \to \Sigma^{-}p$ \\
$\Sigma^{-}p \to \Lambda n$ inelastic scattering & $\Sigma^{-}p \to \Lambda n \to (p \pi^{-}) n$\\
$\Sigma^{-}p \to \Sigma^{0} n$ inelastic scattering & $\Sigma^{-}p \to \Sigma^{0} n  \to \Lambda \gamma n \to (p \pi^{-}) \gamma n$\\ 
\hline\hline 
Scattering of the $\Sigma^{-}$ decay products with proton&   \\
\hline\hline 
$np$ scattering & $\Sigma^{-} \to n \pi^{-}$ followed by $np \to np$ \\
$\pi^{-}p$ scattering & $\Sigma^{-} \to n \pi^{-}$ followed by $\pi^{-}p \to \pi^{-}p$ \\
\hline\hline
\end{tabular}
\end{center}
\end{table*}

Next, we explain how the scattering events can be identified by considering an example of the $\Sigma^{-}p$ elastic scattering.
The momentum vector of the $\Sigma^{-}$ particle is reconstructed from the spectrometer information.
The recoil angle of proton is obtained as the crossing angle between the $\Sigma^{-}$ and  proton tracks.
The proton's energy can be calculated from the recoil angle of the proton by applying  $\Sigma^{-}p$ elastic scattering kinematics.
Here, the calculated energy is described as $E_{calculated}$.
On the other hand, the proton energy was measured by the CFT and the BGO and  is denoted as $E_{measured}$ here.
We then define $\Delta E (\Sigma^{-}p)$ as the difference between $E_{measured}$ and $E_{calculated}$, that is, $\Delta E (\Sigma^{-}p) = E_{measured}-E_{calculated}$, under the $\Sigma^{-}p$ elastic scattering assumption.
If the recoil proton  really originated in the $\Sigma^{-} p$ elastic scattering, such events would make a peak around $\Delta E(\Sigma^{-}p) = 0$ in the $\Delta E(\Sigma^{-}p)$ spectrum.
We checked the kinematical consistencies for all possible secondary reactions.
These are the $np, \pi^{-}p,  \Sigma^{-}p$  elastic scatterings and $\Sigma^{-}p \to \Lambda n$ reaction in the same manner as described in the $\Sigma^{-}p$ scattering case  except for the $\Sigma^{-}p \to \Sigma^{0}n \to \Lambda \gamma n$ reaction where kinematical reconstruction was impossible owing to the missing information of $\gamma$ from the $\Sigma^{0}$ decay.
In these calculations, the $\pi^{-}$ momentum  was determined by imposing some assumptions corresponding to each reaction.
In the $np$ scattering, a similar value of the $\Delta E (np)$ was defined as the difference between $E_{measure}$ and $E_{calculated}$ calculated from the recoil angle based on the $np$ scattering kinematics, where the initial neutron momentum was determined by assuming that $\pi^{-}$ was emitted in the $\Sigma^{-}$ decay.
Similarly, in the case of the $\Sigma^{-}p \to \Lambda n$ reaction assumption, $\Delta p(\Sigma^{-}p \to \Lambda n)$ was defined as the difference between the $\Lambda$ momentum calculated from the scattering angle and  that reconstructed from the $\Lambda$ decay where the momentum magnitude of $\pi^{-}$ was determined so that the invariant mass of $\pi^{-}$ and proton became the $\Lambda$ mass.
In the $\pi^{-}p$ scattering, a similar momentum difference $\Delta p (\pi^{-}p)$ was obtained for the recoil proton.

\begin{figure*}[]
\begin{center}
\begin{minipage}{7.5cm}
 \includegraphics[width=0.98\textwidth]{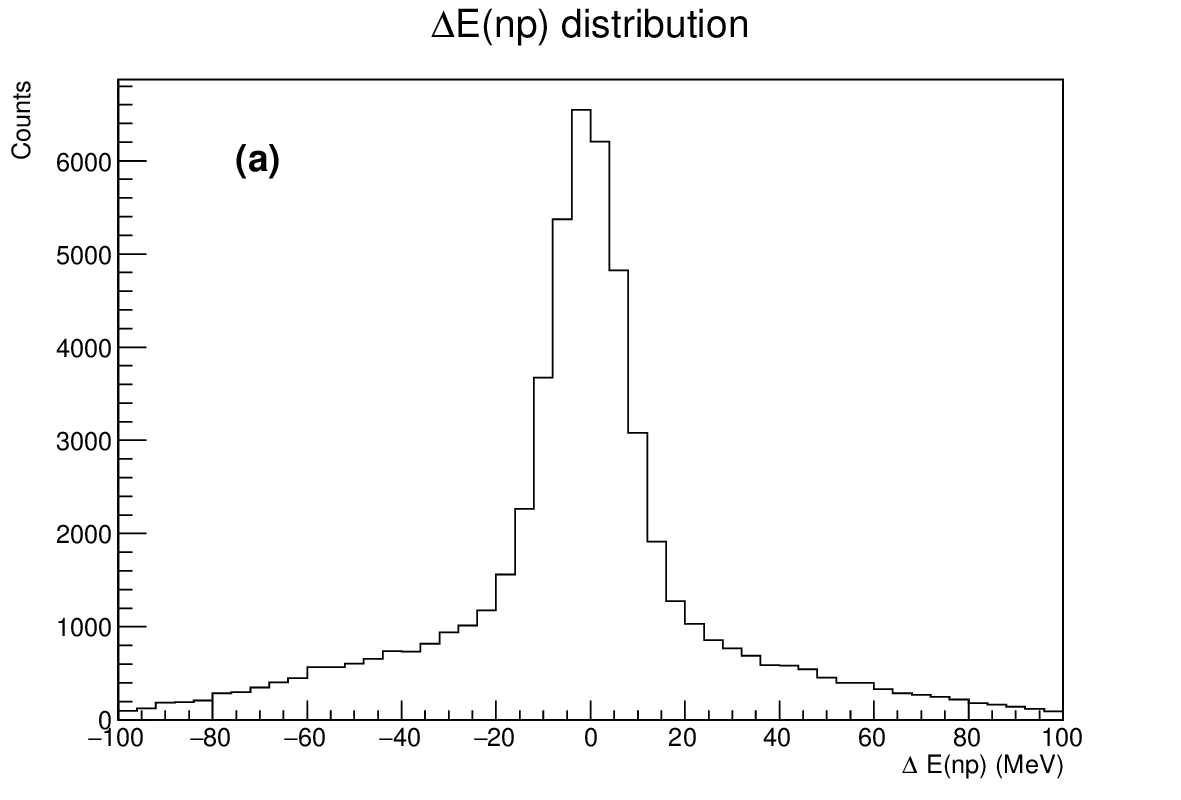}
\end {minipage}
\hspace{0.2cm}
\begin{minipage}{7.5cm}
 \includegraphics[width=0.98\textwidth]{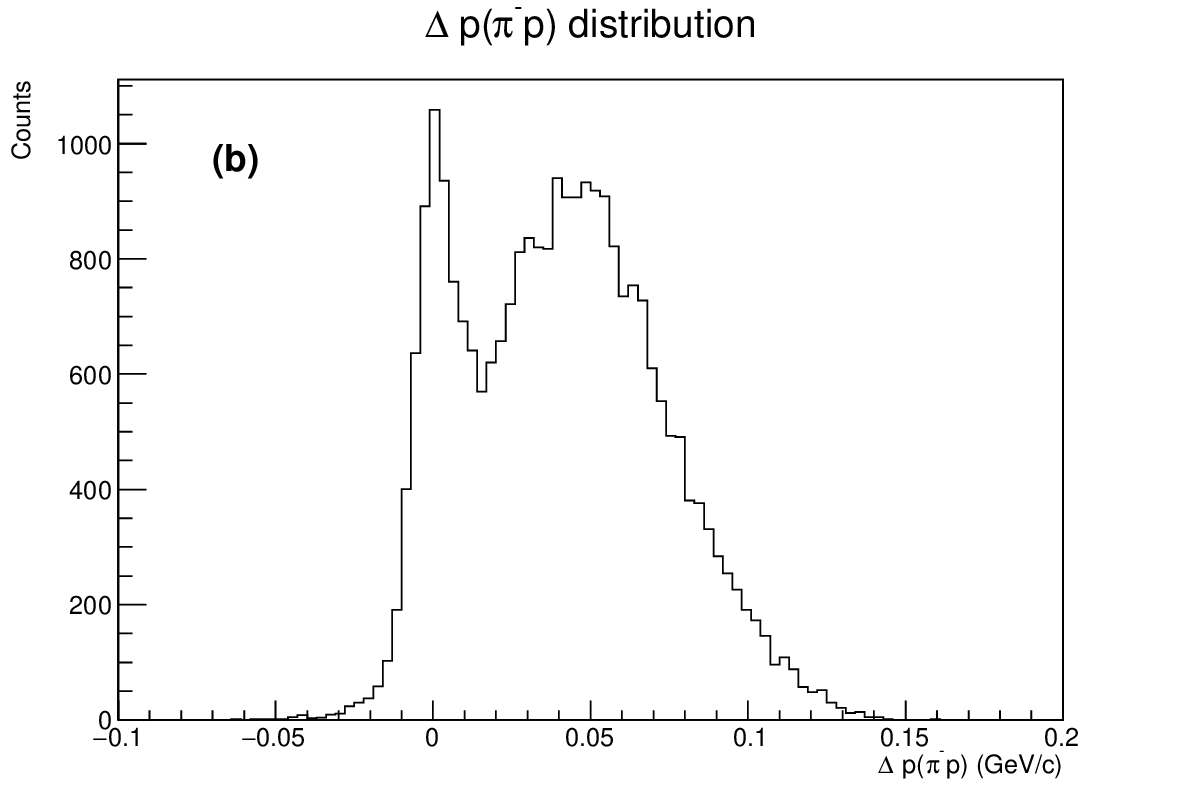}
 \end{minipage}
 
 \begin{minipage}{7.5cm}
 \includegraphics[width=0.98\textwidth]{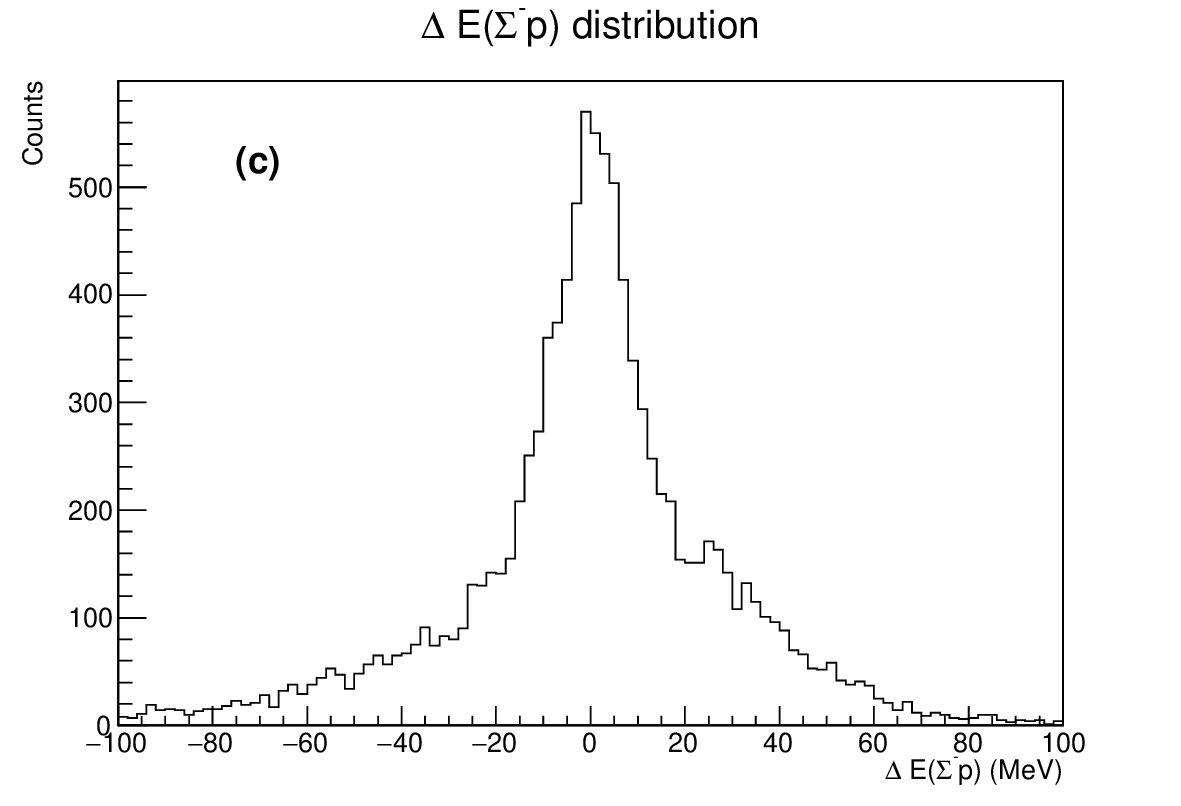}
\end {minipage}
\hspace{0.2cm}
\begin{minipage}{7.5cm}
 \includegraphics[width=0.98\textwidth]{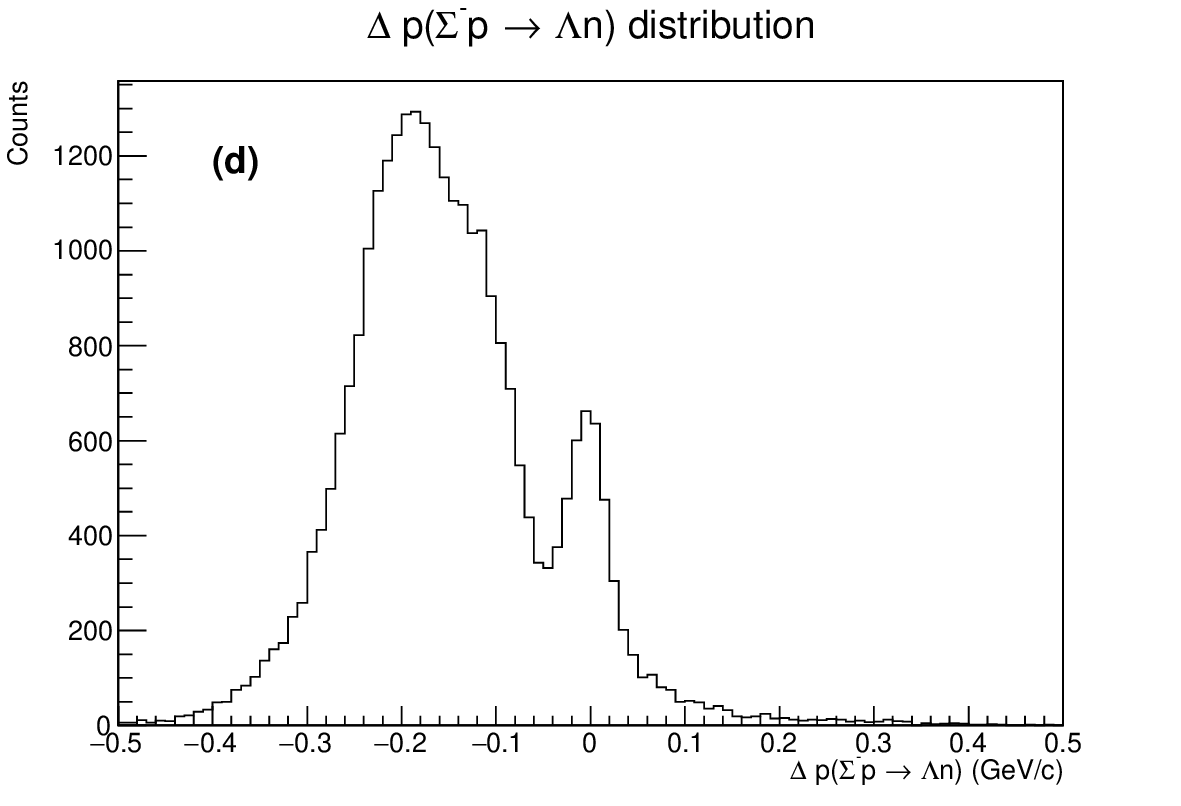}
 \end{minipage}

   \caption{ Kinematical consistencies between the measured  and calculated energies from the scattering angle assuming the four different scattering reactions.
   (a) The $\Delta E (np)$ distribution, 
   (b)  $\Delta p (\pi^{-}p)$ distribution,
   (c)  $\Delta E (\Sigma^{-}p)$ distribution,  and
   (d)  $\Delta p (\Sigma^{-}p \to \Lambda n)$ distribution.
Peaks around $\Delta E=0$ or $\Delta p =0$ correspond to the assumed reaction events.
   }
      \label{showDeltaE_NpScat}
   \end{center}
\end{figure*}

Fig. \ref{showDeltaE_NpScat} (a), (b), (c) and (d) show the results for the assumption of the $np$ and  $\pi^{-}p$ scatterings after the $\Sigma^{-}$ decay,  the $\Sigma^{-}p$ elastic scattering, and the $\Sigma^{-}p \to \Lambda n$ reaction, respectively.
The peaks  around  $\Delta E=0$ or $\Delta p=0$  correspond to the assumed reaction events, as $\Delta E(\Sigma^{-}p)$ for the $\Sigma^{-}p$ scattering events is shown  in  Fig. \ref{showDeltaE_NpScat} (c) for example.
In deriving the $\Delta E (\Sigma^{-}p)$ spectrum, we applied the following analysis cuts to obtain a better signal to noise (S/N) ratio.
We required that the scattering vertex and decay vertex of the scattered $\Sigma^{-}$, which were defined as the closest points between the incident $\Sigma^{-}$ particle and  recoil proton and between the scattered $\Sigma^{-}$ and  decay $\pi^{-}$, respectively, should be within 30 mm from the target center in the $xy$  plane.
The closest distances at the two vertices were required to be less than 12 mm and 15 mm, respectively.
The relative distance in the $z$ vertex positions for these two vertices defined as $\Delta(vtz) = vtz_{decay} - vtz_{scattering}$ was required to be $-10 < \Delta(vtz)$ (mm) $<50$ considering the lifetime of $\Sigma^{-}$.
To improve the S/N ratio in the $\Delta E(\Sigma^{-}p)$ spectrum,  events around the peak regions of the $\Delta E(np)$, $\Delta p(\pi^{-}p)$ and $\Delta p(\Sigma^{-}p \to \Lambda n)$ spectra were removed.
However, the contamination ratio of these background events depends on the $\Sigma^{-}$ beam momentum and scattering angle.
Therefore, the cuts for  $\Delta E(np)$, $\Delta p(\pi^{-}p)$ and $\Delta p(\Sigma^{-}p \to \Lambda n)$ were optimized for each scattering angle and each $\Sigma^{-}$ beam momentum.
Finally, approximately 4,500 events have been observed for the $\Sigma^{-}p$ elastic scattering from 1.62$\times 10^{7}$ $\Sigma^{-}$ particles.

\section{Analysis 2 : Derivation of the differential cross section of the $\Sigma^{-}p$ elastic scattering}\label{sec_ana2}
The differential cross section was defined as

\begin{equation}
\frac{d \sigma}{d \Omega} = \frac{\sum_{i_{vtz}} \frac{N_{scat}(i_{vtz}, \cos \theta)}{\epsilon(i_{vtz}, \cos \theta)}}{\rho \cdot N_{A} \cdot L \cdot \Delta \Omega}, \label{equ_cs}
\end{equation}
where $\rho$, $N_{A}$ and  $L$ represent the target density, Avogadoro's number, and  total flight length of the $\Sigma^{-}$ hyperons in the LH$_{2}$ target, respectively.
The values of the numerator depend on the vertex position of the $\Sigma^{-}p$ scattering.
$i_{vtz}$ represents the index of the $z$ vertex position from $-150$ mm to 150 mm with an interval of 30 mm.
$N_{scat}(i_{vtz}, \cos \theta)$  and $\epsilon(i_{vtz}, \cos \theta)$ represent the number of  $\Sigma^{-}p$ elastic scattering events and the detection efficiency of CATCH for the scattering angle $\theta$ in the c.m. frame for  the $z$ vertex position of $i_{vtz}$, respectively.
$\Delta \Omega$ represents the solid angle for each scattering angle.
The determination of these values is described in the following subsection.
The differential cross section is finally obtained.

\subsection{Total flight length of the $\Sigma^{-}$ beam in the LH$_{2}$ target}

\begin{figure}[t]
\begin{center}
\begin{minipage}{7.5cm}
 \includegraphics[width=0.98\textwidth]{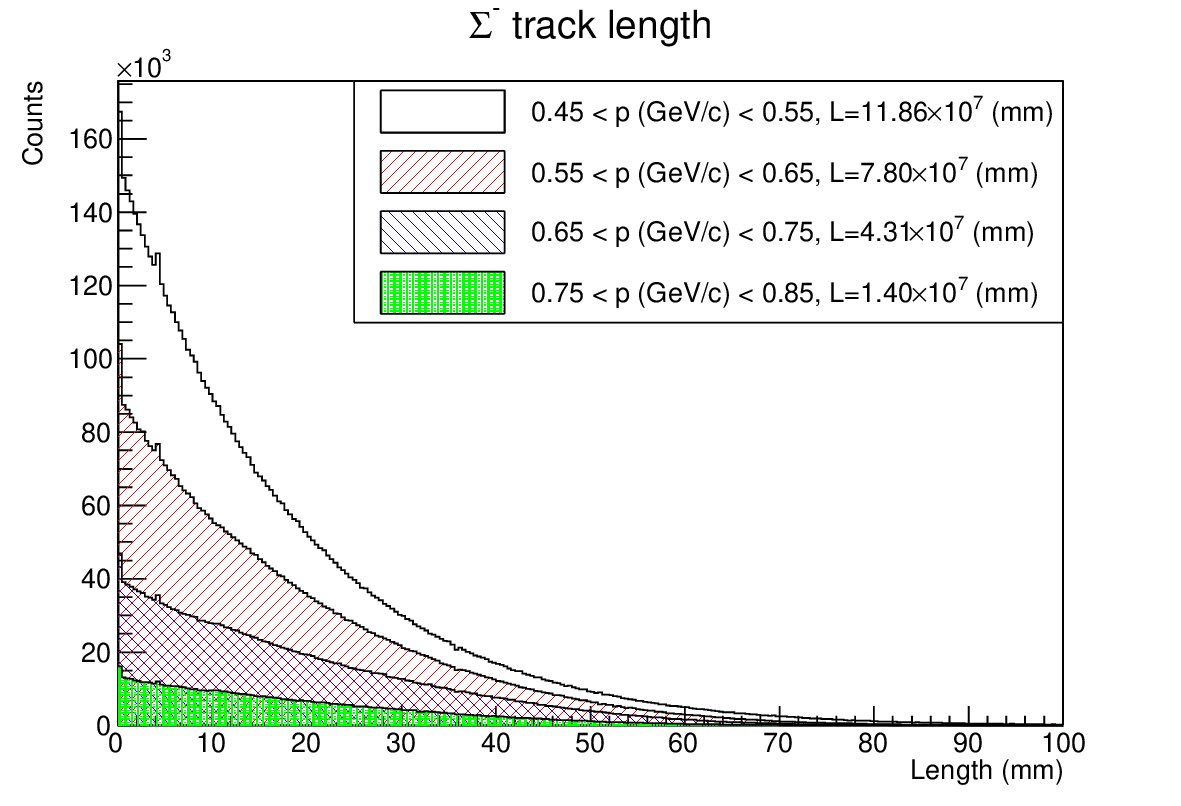}
\end {minipage}
\hspace{0.2cm}
\begin{minipage}{7.5cm}
 \includegraphics[width=0.98\textwidth]{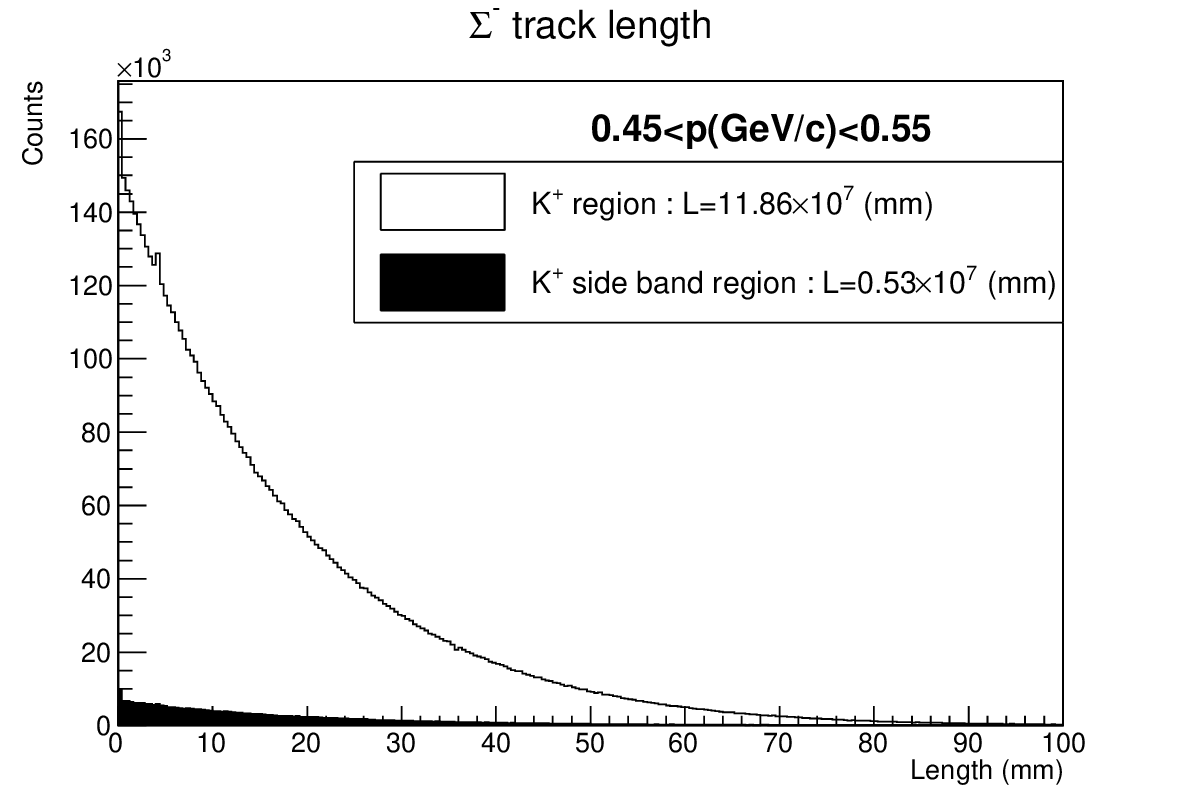}
   \end{minipage}
   \caption{ (Top) Estimated $\Sigma^{-}$ track length in the LH$_{2}$ target for the four momentum regions.
   (Bottom) Contribution in the $\Sigma^{-}$ track length of the miscalculated event  in  the $m^{2}$ distribution under the $K^{+}$ peak (filled histogram).
   Total track lengths ($L$) are also shown by summing up each track length.
   }
      \label{showSigmaTrackLengthForEps_c1}
   \end{center}
\end{figure}

Because the $\Sigma^{-}$ particles decay inside the LH$_{2}$ target, a standard expression of $\rho \cdot N_{A} \cdot t \cdot N_{beam}$ , where $t$ and $N_{beam}$ represent the target thickness and the number of beam particles, respectively, cannot be used.
Instead, we estimate the total flight length of $\Sigma^{-}$ in the LH$_{2}$ target  ($L$).

The $\Sigma^{-}$ flight length was not measured event by event in this experiment.
However, the $\Sigma^{-}$ momentum and the production point were obtained from the spectrometer analysis.
The total flight length ($L$) could be obtained from a Monte Carlo simulation considering the $\Sigma^{-}$ lifetime using the obtained momentum and vertex information of $\Sigma^{-}$ from the spectrometer analysis as the inputs of the simulation.
In this simulation, the energy deposit of $\Sigma^{-}$ was also considered.
Fig. \ref{showSigmaTrackLengthForEps_c1} (top) shows the estimated $\Sigma^{-}$ flight length  in the LH$_{2}$ target for each event  for the four momentum regions.
The total flight lengths ($L$)  were obtained by adding all flight lengths in each event and are shown in the legend in the histogram, that is, $10^{8} \sim 10^{7}$  mm  in total.
As shown in Fig. \ref{showSigmaWithSB} (a),  contamination events occured in the selected $\Sigma^{-}$ mass region owing to the misidentification of $K^{+}$.
The filled histogram in Fig. \ref{showSigmaTrackLengthForEps_c1} (bottom) shows the flight length by applying the $\Sigma^{-}$ production kinematics for the side-band events.
This contaminated flight length was subtracted from the total flight length in driving the differential cross section.

\color{black}
If the $\Sigma^{-}$ momentum obtained from the missing momentum of the $\pi^{-}p \to K^{+}X$ reaction was systematically shifted, the total flight length would be affected.
To estimate the accuracy of the missing momentum, the $\pi^{-}p$ elastic scattering between the $\pi^{-}$ beam and a proton in the LH$_{2}$ target was used where the scattered $\pi^{-}$ and recoil proton were detected by KURAMA and CATCH, respectively.
The proton's momentum obtained from the missing momentum of the  $\pi^{-}p \to \pi^{-}X$ reaction was compared with the one measured by CATCH.
From this study, the systematic difference for the $\Sigma^{-}$ momentum was estimated as 4 MeV/$c$ at maximum.
The systematic uncertainty for the $\Sigma^{-}$ total flight length is less than 1\%, which is much smaller than other uncertainties.
The contribution of the uncertainty of the total flight length in the differential cross section is summarized in Table \ref{table_dsdw1}, \ref{table_dsdw2}, \ref{table_dsdw3} and \ref{table_dsdw4} in Appendix \ref{appendix_2}. 
\color{black}

\subsection{Detection efficiency in CATCH}
\begin{figure}[t]
\begin{center}
\begin{minipage}{7.5cm}
 \includegraphics[width=0.98\textwidth]{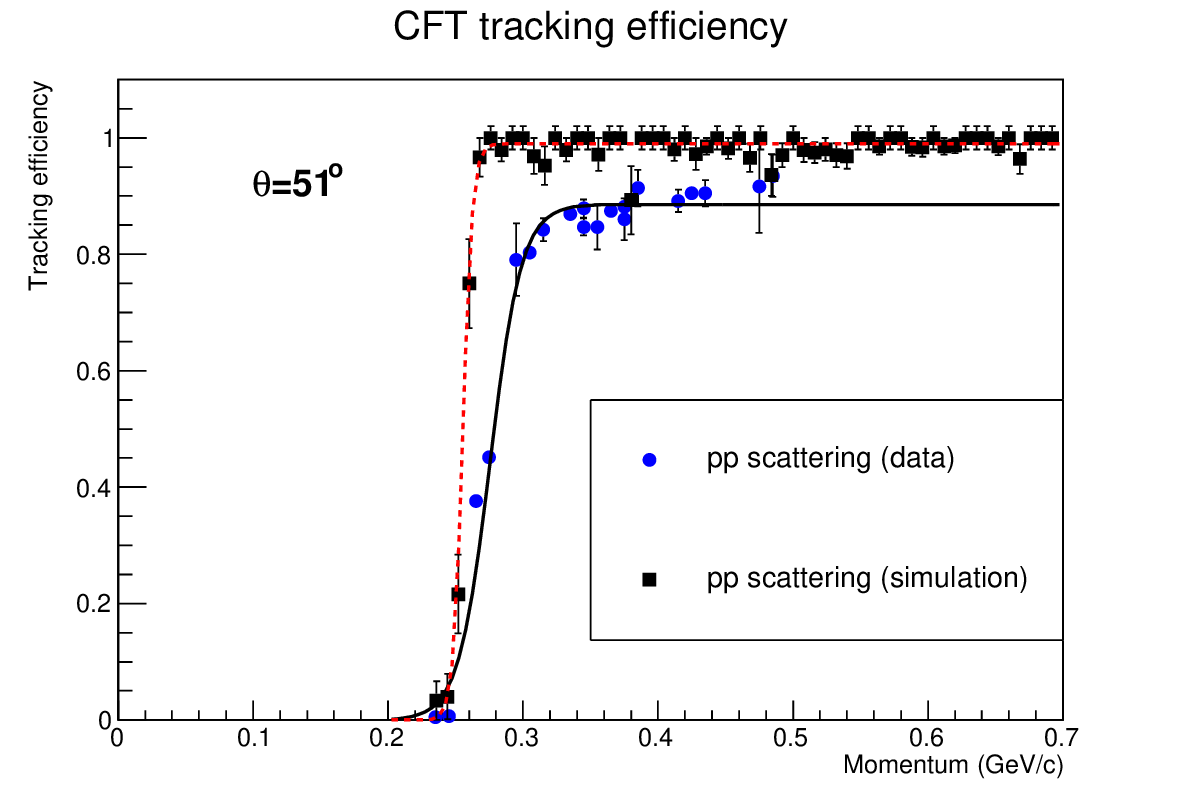}
\end {minipage}
\hspace{0.2cm}
\begin{minipage}{7.5cm}
 \includegraphics[width=0.98\textwidth]{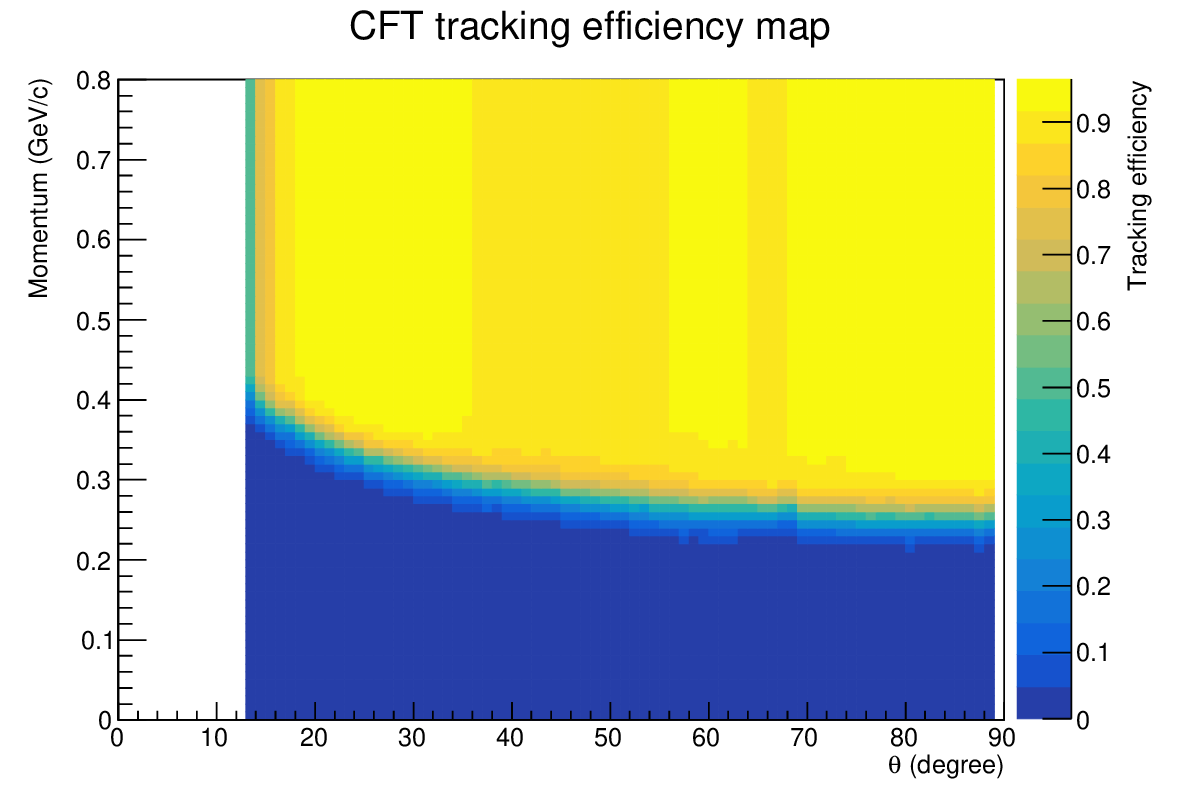}
   \end{minipage}
   \caption{ (Top) Momentum dependence of the CFT tracking for the proton obtained from the $pp$ scattering data and its simulation for a scattering angle of 51$^{\rm o}$ in the laboratory frame.
   (Bottom) Estimated CFT tracking efficiency map as a function of the scattering angle and  momentum of the proton.
   }
      \label{calcTrackEffMomDepForEachTheta_c7}
   \end{center}
\end{figure}

The detection efficiency in CATCH includes the detector acceptance, the tracking efficiency of the CFT, and the energy measurement efficiency of the BGO.
These depend on the angle and  momentum of a particle.
Particularly, the efficiency of a proton significantly depends on the momentum because the energy loss in the materials in the LH$_{2}$ target system and the fibers in the CFT increases for lower momentum.
To obtain a realistic efficiency, we used  the $pp$ scattering data where two protons were emitted.
By identifying the $pp$ scattering event by detecting one proton and checking the kinematics between the scattering angle and kinetic energy of the proton, the other proton's angle and momentum could be predicted from the missing momentum of the $pp \to pX$ reaction.
The CFT tracking efficiency was obtained by checking whether the predicted track could be detected or not.
Using the $pp$ scattering data with different incident beam momenta from 450 to 850 MeV/$c$, the momentum dependence of the recoil proton can be obtained.
Fig. \ref{calcTrackEffMomDepForEachTheta_c7}  (top) shows the momentum dependence of the measured tracking efficiency  for the protons emitted at $\theta=51^{\rm o}$ in the laboratory frame.
In Fig. \ref{calcTrackEffMomDepForEachTheta_c7} (top), the result for the simulation based on the Geant4 package \cite{Agostinelli:2003} is also shown.
The realistic efficiency estimated from the $pp$ scattering data is slightly lower than that in the simulation, and the efficiency decreases from the slightly higher momentum compared with the simulation with a gradual slope.
The lower efficiency was attributed to  a gap between fibers in the CFT and the difference at the lower momentum region indicated that the realistic amount of material in the experimental setup was larger than that considered in the simulation.
The efficiency was modeled as the Fermi function with three parameters, that is, the maximum efficiency ($P_{max}$), the momentum with half efficiency ($P(1/2)$) and blurriness ($\mu$).
These parameters were estimated  by fitting the momentum dependence of the efficiency in the $pp$ scattering data as shown in the solid line in Fig.  \ref{calcTrackEffMomDepForEachTheta_c7} (top).
The parametrization of the CFT tracking efficiency was performed for each scattering angle  using the $pp$ scattering data.
However, there was an uncertainty resulting from the momentum calibration of the recoil proton and so on.
We searched for possible parameter values to explain the differential cross sections of the $pp$ scattering.
This uncertainty was considered for the CATCH efficiency estimation for the $\Sigma^{-}p$ scattering described in subsection \ref{sec_eff_SMp}.
Fig. \ref{calcTrackEffMomDepForEachTheta_c7} (right) shows the obtained efficiency map for the CFT tracking.
The detector acceptance of CFT is also included in this efficiency map.

\begin{figure}[t]
\begin{center}
\begin{minipage}{7.5cm}
 \includegraphics[width=0.98\textwidth]{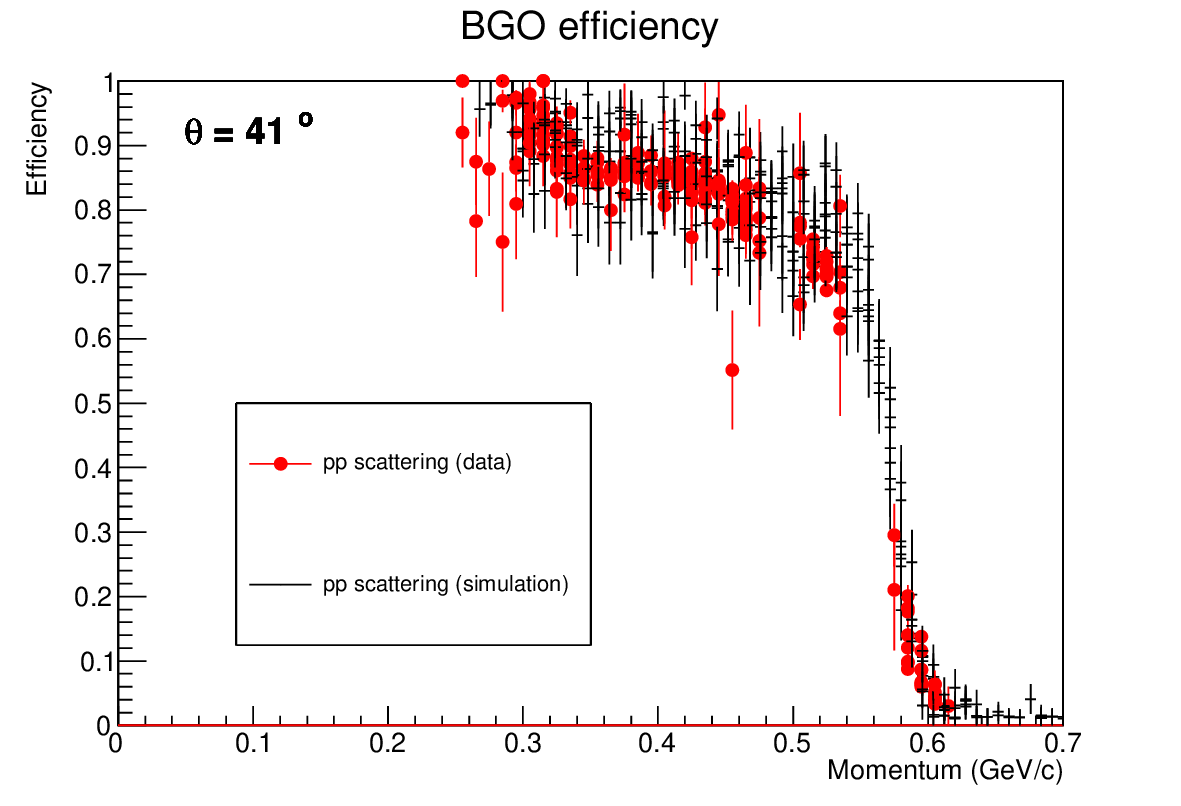}
\end {minipage}
\hspace{0.2cm}
\begin{minipage}{7.5cm}
 \includegraphics[width=0.98\textwidth]{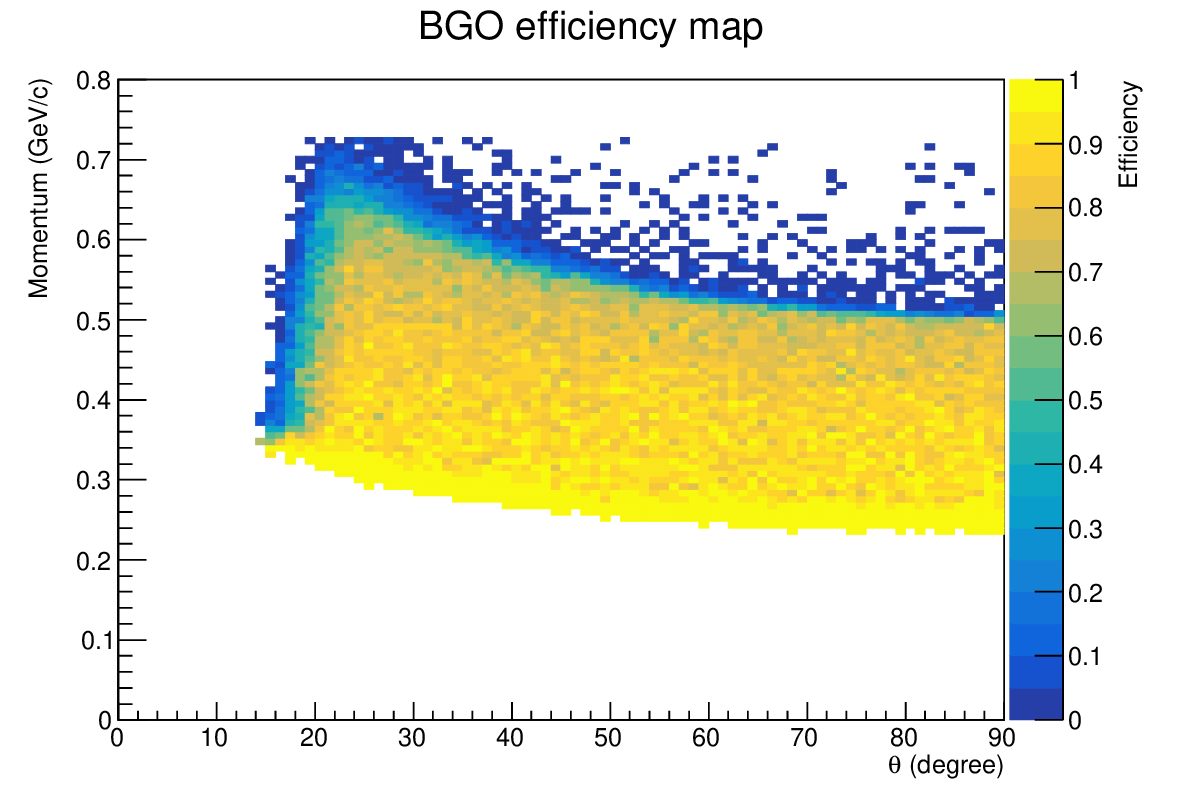}
   \end{minipage}
   \caption{ (Top) Momentum dependence of the BGO efficiency for the proton obtained from the $pp$ scattering data for a scattering angle of 41$^{\rm o}$ in the laboratory frame.
   The red and cross points show the estimated efficiencies from the $pp$ scattering data and Geant4 simulation.
   (Bottom) Estimated BGO efficiency map as a function of the scattering angle and momentum of the proton.
   }
      \label{showEffMomDepForEachTheta_wData_2019_2020_separate_ForBGOeps_c5}
   \end{center}
\end{figure}

Next, the efficiency of BGO was estimated using the $pp$ scattering.
In this case, we checked whether or not the measured energy by BGO was consistent with the predicted energy from the $pp$ scattering kinematics.
Fig. \ref{showEffMomDepForEachTheta_wData_2019_2020_separate_ForBGOeps_c5} (top) shows the comparison of the momentum dependence of the BGO efficiency  for protons emitted at $\theta=41^{\rm o}$ in the laboratory frame for both the $pp$ scattering data (circular points) and  simulation (crossed lines).
In this efficiency, the acceptance of BGO was also included.
For the BGO efficiency, the consistency between the data and simulation was obtained effectively for all the angular regions.
Therefore, the simulated efficiency was used as the efficiency map as shown in Fig. \ref{showEffMomDepForEachTheta_wData_2019_2020_separate_ForBGOeps_c5} (bottom) .

\subsection{Detection efficiency for the $\Sigma^{-}p$ scattering events} \label{sec_eff_SMp}

\begin{figure}[]
  \centerline{\includegraphics[width=0.5\textwidth]{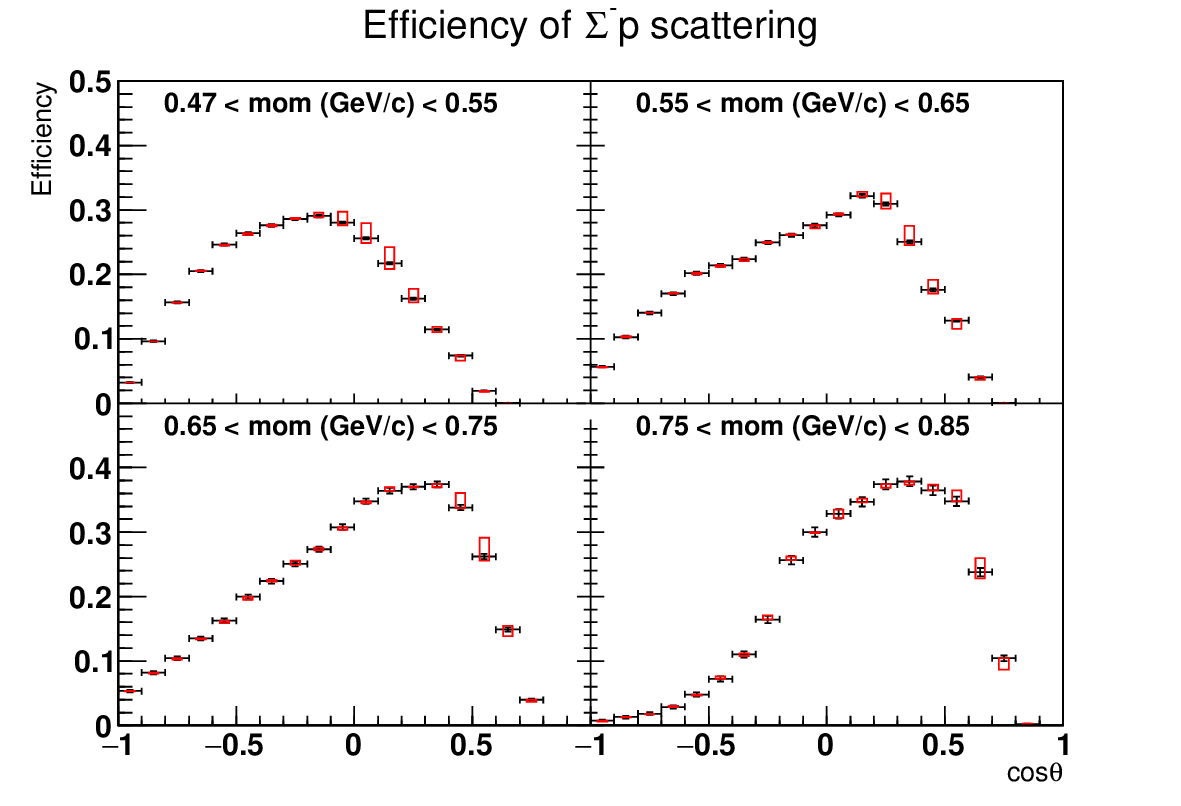}}
  \caption{ Detection efficiency for the $\Sigma^{-}p$ elastic scattering events for the four $\Sigma^{-}$ beam momentum regions.
  These efficiencies were the averaged values for all the vertex regions of $-150 < vtz$ (mm) $< 150$.
  }
  \label{compSigmaMP_Eff_c4}
\end{figure}

The detection efficiency for the  $\Sigma^{-}p$ scattering events ($\epsilon(i_{vtz}, \cos \theta)$) was estimated based on the previous study of the CATCH efficiency.
We generated the $\Sigma^{-}p$ scattering event in the Monte Carlo simulation.
In this simulation, not only the $\Sigma^{-}p$ elastic scattering but also other channels including the $\Sigma^{-}p \to \Lambda n$ and $\Sigma^{0}n$ reactions and secondary scatterings of the $\Sigma^{-}$ decay products such as the $np$ and the $\pi^{-}p$ scatterings were included to study the background contribution in the $\Delta E$ spectra for the kinematical identification.
Here, the secondary reactions of $np$ and $\pi^{-}p$ are generated based on the cross sections of theoretical calculations \cite{Fujiwara:2007} and past measurements \cite{PDG:2020}.
The three $\Sigma^{-}p$ reactions were assumed to be 2.4 mb/sr with an isotropic distribution in the c.m. system in the simulation.
It is important that the kinematical distributions from the background contributions can be studied from this simulation.
These generated data were analyzed by the same analysis program considering the realistic angular and vertex resolutions of CFT and the spectrometers.
The energy resolution of BGO is also included based on the measured resolution expressed in equation (\ref{eq_bgo_reso}).
To identify the $\Sigma^{-}p$ scattering, one proton and one $\pi^{-}$ should be detected by CATCH.
In this study, the estimated efficiency of the CFT tracking and BGO efficiency were used for the proton.
For $\pi^{-}$, the CFT tracking efficiency was estimated using the tracking efficiency with the same velocity ($\beta$) obtained from the proton results.
The same analysis cuts described in Section \ref{sec_ana_sigmaMp} were applied to the simulated data to estimate the efficiency of the analysis cuts.
Fig. \ref{compSigmaMP_Eff_c4} shows the detection efficiency as a function of the  scattering angle in the c.m. frame ($\cos \theta$) for the four $\Sigma^{-}$ beam momentum regions.
These efficiencies were the averaged values for  all the vertex regions of $-150 < vtz$ (mm) $< 150$.
In the derivation of the differential cross section, the efficiency for each $z$ vertex was also used to take into account the vertex dependence of the efficiency.
Generally, the efficiency decreases for the forward scattering event near $\cos \theta = 1$, where  $\Sigma^{-}$ is scattered at the forward angle and the proton is recoiled  backward with a low energy.
The low energy proton can not be detected and this results in lower efficiency.
For the backward angle around $\cos \theta = -1$ , the situation is opposite where the proton is recoiled  forward with a high energy.
However, because of the acceptance of CATCH, the efficiency decreases.
At the higher  $\Sigma^{-}$ momentum over 750 MeV/$c$, the recoil proton energy is too high to be stopped in the BGO.
Therefore the tendency of the efficiency less than $\cos \theta = -0.4$ is different compared with the other lower beam momentum region.

\color{black}
The CFT tracking efficiency was estimated using the reasonable modeling described in the previous subsection.
However, there was an uncertainty resulting from the momentum calibration of the recoil proton and so on.
We changed the $P_{1/2}$ and $\mu$ within the possible region where the differential cross sections of the $pp$ scattering were derived as reasonable values.
This uncertainty is relatively large for a proton with a small momentum less than 300 MeV/$c$.
This uncertainty was considered for the CATCH efficiency estimation for the $\Sigma^{-}p$ scattering.
The red box in Fig. \ref{compSigmaMP_Eff_c4} shows the uncertainty attributing from the uncertainty of the CFT tracking efficiency for the low momentum region.
Therefore, the uncertainty increases at the acceptance edge of the forward angle owing to the low energy of the recoil proton.
To check the validity of the derived efficiency, the differential cross sections of the $np$ scattering from the $\Sigma^{-}$ decay were also derived from the peak counts in the $\Delta E(np)$ spectra based on the efficiency of the $np$ scattering event made by the same method.
These values were consistent with past data within 3$\sigma$.
\color{black}

\subsection{Detection number of the $\Sigma^{-}p $ scattering events for each scattering angle}

\begin{figure}[]
\begin{center}
 %
  \centerline{\includegraphics[width=0.5\textwidth]{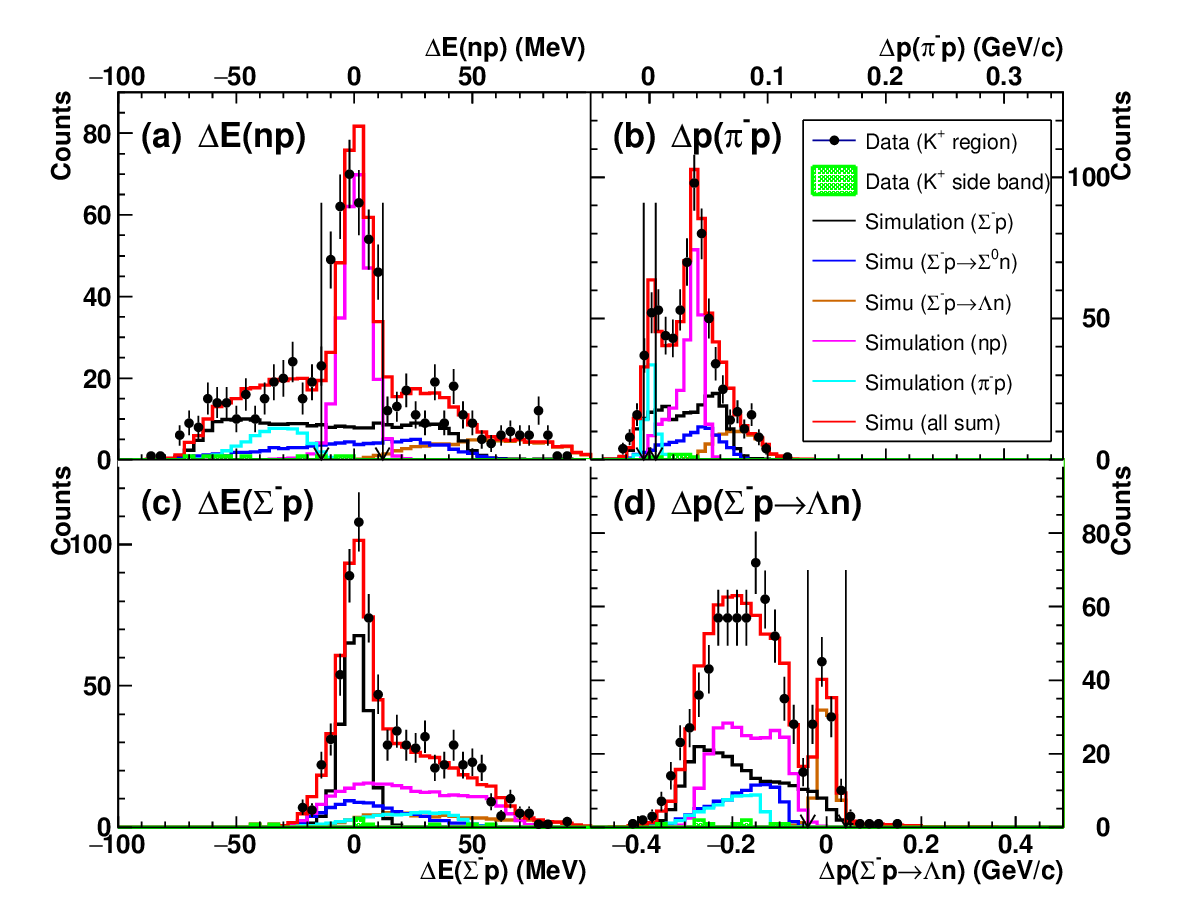}}
   \caption{ Kinematical consistency between the measured and  calculated energies from the measured scattering angle assuming the four different scattering processes,
   (a) $\Delta E (np)$, 
   (b) $\Delta p (\pi^{-}p)$,
   (c) $\Delta E (\Sigma^{-}p)$, and 
   (d) $\Delta p (\Sigma^{-}p \to \Lambda n)$ distributions, 
    for the angular region of $0 \le \cos \theta \le 0.1$ in the momentum range of 470 $<$ $p$ (MeV/$c$) $<$ 550.
Data points with error bars and a green shaded histogram show the experimental data for the $K^{+}$ region and the side-band region of $K^{+}$ in the mass square spectrum, respectively.
Simulated spectra for the assumed reactions are also shown and the histogram with a red line shows the sum of these spectra.
The arrows in (a), (b), and (d) show the cut regions for the background suppression.
   }
      \label{showSigmaMPScatWithFit4_wBG_result_0_c1}
   \end{center}
\end{figure}

The number of the $\Sigma^{-}p$ elastic scattering events was estimated from the peak counts in the $\Delta E (\Sigma^{-}p)$ spectra for each scattering angle.
The understanding of the background structure in the $\Delta E (\Sigma^{-}p)$ spectra is essential to estimate the number of scattering events.
The sources of the background are listed in Table \ref{relation_secondary_reaction}.
All background reactions except for the $\Sigma^{-}p \to \Sigma^{0}n$ reaction can be identified by making each $\Delta E$ or $\Delta p$ spectrum, which show the kinematical consistency for the assumed reaction.
Fig. \ref{showSigmaMPScatWithFit4_wBG_result_0_c1} shows the four kinematical plots for the $np$, $\pi^{-}p$, $\Sigma^{-}p$, and  $\Sigma^{-}p \to \Lambda n$ reactions with the simulated spectra
for the scattering angle  of $0  \le \cos\theta \le 0.1$ in the momentum range of 470 $< p $ (MeV/$c$) $< $ 550. 
In this analysis, the analysis cuts with $\Delta E$ and $\Delta p$ for rejecting the background events are not applied to determine the amounts of the background events from the peak counts in each $\Delta E$ or $\Delta p$ spectrum.
Here, we also took into account the misidentification event of $K^{+}$ by selecting the side-band events of $K^{+}$ in the mass square spectrum.
For example, the $np$ scattering events form the peak structure in Fig. \ref{showSigmaMPScatWithFit4_wBG_result_0_c1} (a) $\Delta E(np)$, whereas they create the background structure for other spectra.
The number of background events, that is, the scale factors of the simulated spectra,  can be effectively estimated by fitting the four kinematical spectra simultaneously, because the contribution of the $np, \pi^{-}p$ and $\Sigma^{-}p \to \Lambda n$ reactions are constrained from the peak counts in each $\Delta E$ or $\Delta p$ spectrum.
All $\Delta E$ spectra are well reproduced by the sum of assumed reactions.

\begin{figure}[]
\begin{center}
 \includegraphics[width=0.5\textwidth]{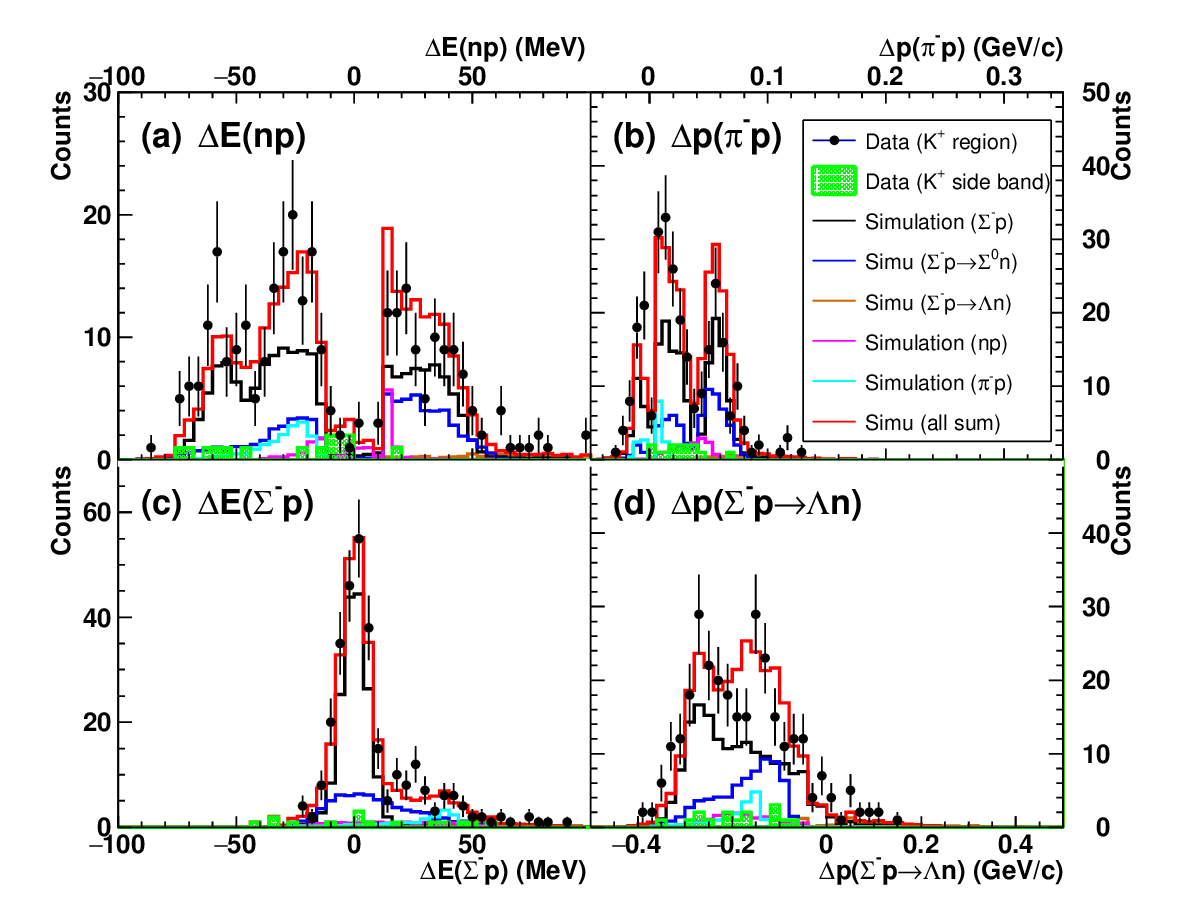}

   \caption{ The same spectra in Fig. \ref{showSigmaMPScatWithFit4_wBG_result_0_c1} after applying a cut to reject the background events by removing the background peak structures in the $\Delta E(np), \Delta p (\pi^{-}p)$ and $\Delta p(\Sigma^{-}p \to \Lambda n)$.
   }
      \label{showSigmaMPScatWithFit4_result_0_c1}
   \end{center}
\end{figure}

In the final analysis, to obtain a better  S/N ratio in the $\Delta E (\Sigma^{-}p)$ spectrum,  
events around the peak region for the background $\Delta E$ spectra are removed if the background contributions are sizable compared to the $\Sigma^{-}p$ event.
As a typical example, the arrows in Fig. \ref{showSigmaMPScatWithFit4_wBG_result_0_c1} show the cut regions for the angular region.
Fig. \ref{showSigmaMPScatWithFit4_result_0_c1} shows the $\Delta E$ spectra after this cut.
These spectra can be fitted again with the simulated spectra with almost the same scale factors obtained in the fit before the background rejection.
Even after these cuts, all spectra could be well reproduced.
Especially, the $\Delta E (\Sigma^{-}p)$ spectrum was well reproduced with the assumed background structure.
The number of the scattering events was obtained from the reproduced spectrum for the $\Sigma^{-}p$ scattering.
The sum of simulated background reactions was also used as the background spectrum.
By applying the cut to reject the background reactions from $\Delta E$ and $\Delta p$, a portion of the $\Sigma^{-}p$ scattering events was also rejected.
This cut efficiency is also considered in the efficiency estimation shown in Fig. \ref{compSigmaMP_Eff_c4}.
The validity of the estimation of the cut efficiency was checked by confirming that the differential cross sections obtained with and without this cut were consistent with each other within the statistical error.
The angular dependences of the $\Delta E (\Sigma^{-}p)$ for the different four $\Sigma^{-}$ beam momentum regions are shown in
Fig. \ref{showSigmaMPScatWithFit4_result_0_c5} and \ref{showSigmaMPScatWithFit4_result_2_c5} in Appendix \ref{appendix_2}.
The spectra were fitted with the same manner where not only the $\Sigma^{-}p$ spectra but also the background $\Delta E$ spectra were fitted simultaneously.

We also fitted only the  $\Delta E$($\Sigma^{-}p$) spectrum with the simulated spectra to study the systematic uncertainty due to the background estimation.
The differential cross sections were then derived with different background estimations and the difference was included into the systematic error as described in the next subsection.

\subsection{Differential cross sections and comparison with theoretical models}

\begin{figure*}[]
  \centerline{\includegraphics[width=0.6\textwidth]{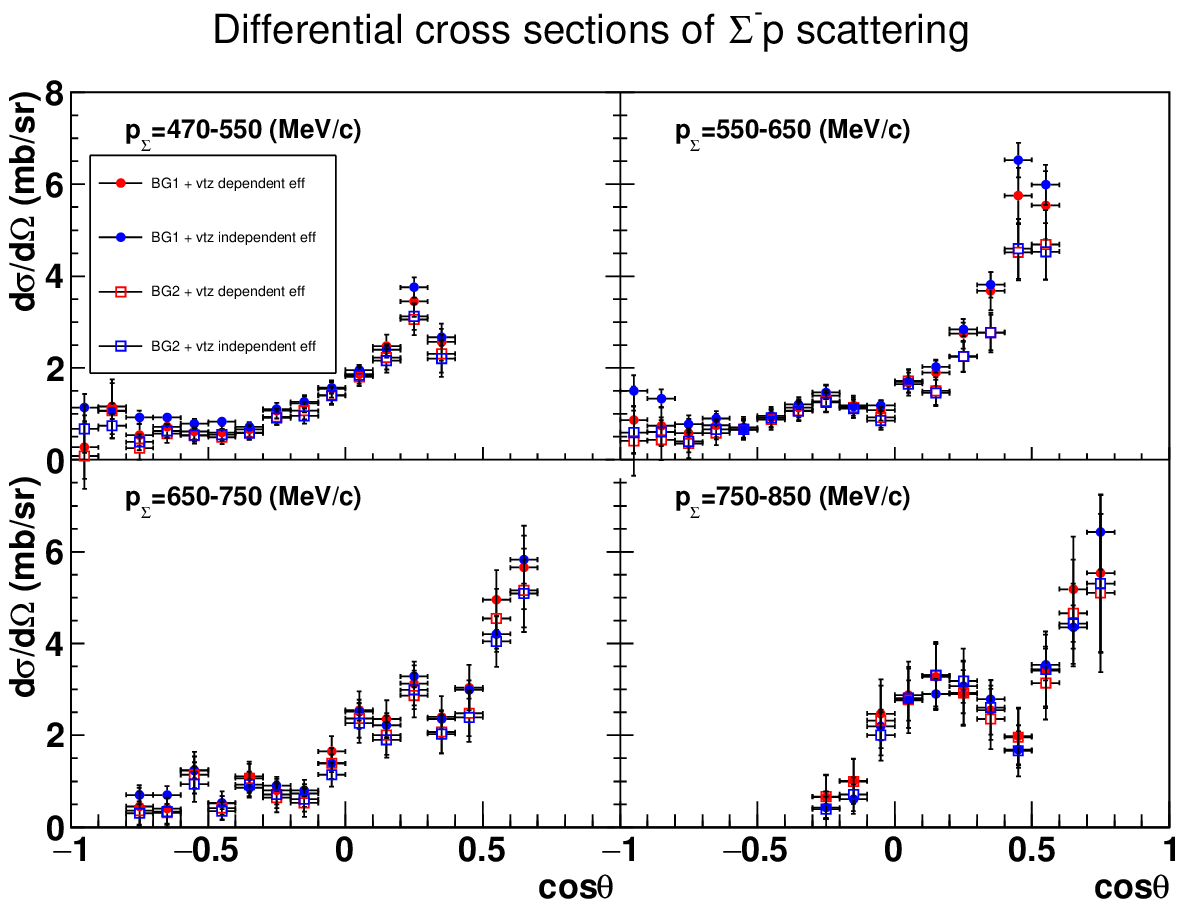}}
  \caption{  Differential cross sections for four momentum regions derived from different methods and background structures.
The closed points and open points were obtained from BG1 obtained from the simultaneous fit for all four $\Delta E, \Delta p$ spectra and BG2 obtained from the single fit of the $\Delta E (\Sigma^{-}p)$, respectively.  
The red and blue colors were obtained by the $z$ vertex dependent  and  $z$ vertex averaged methods, respectively.
  }
  \label{showAllSigmaMPdSdW3_1}
\end{figure*}

\begin{figure*}[]
  \centerline{\includegraphics[width=0.6\textwidth]{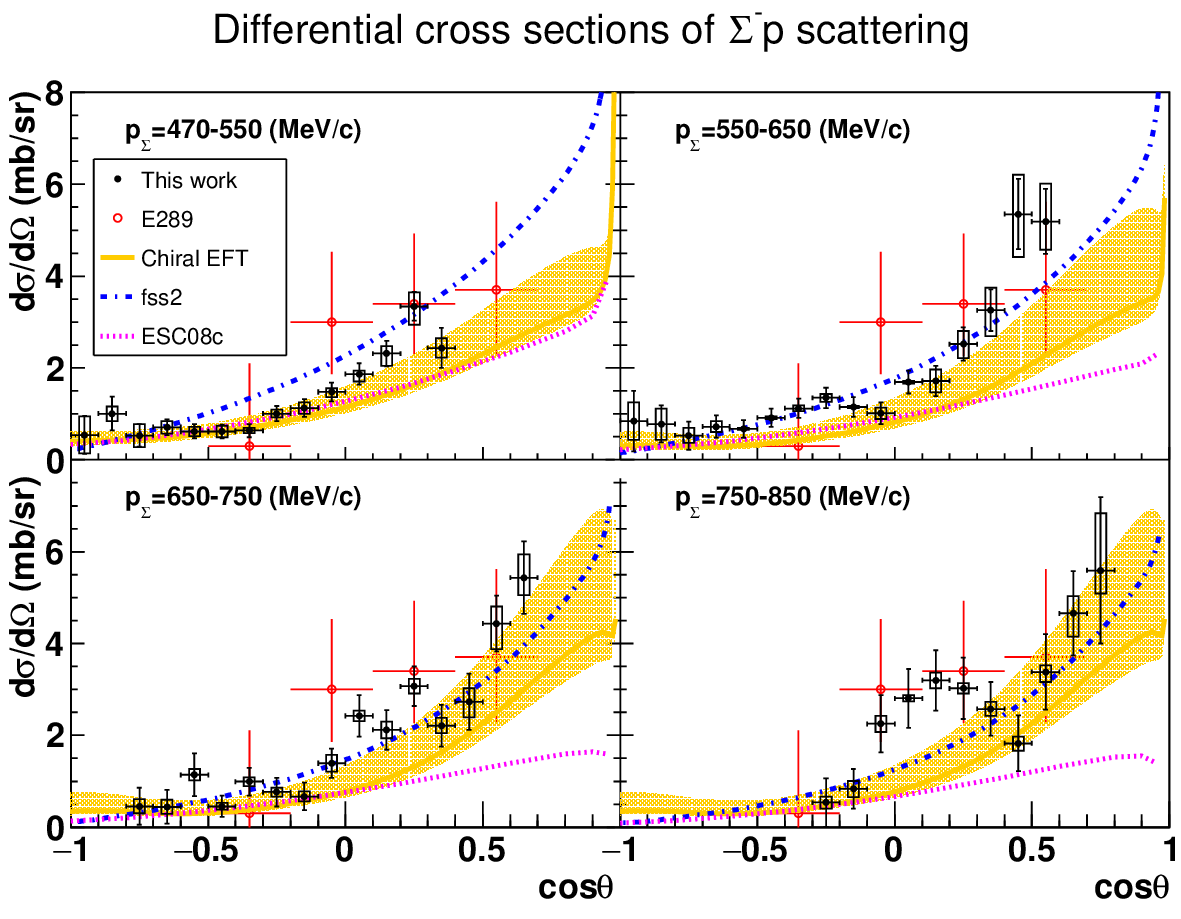}}
  \caption{ Differential cross sections obtained in the present experiment (black points). The error bars and boxes show  statistical and   systematic uncertainties, respectively.
  The red points are averaged differential cross sections of 400 $<$ $p$ (MeV/$c$) $<$ 700 taken in KEK-PS E289 (the same points are plotted in all of the four momentum regions).
  The dotted (magenta), dot-dashed (blue) and solid (yellow) lines represent the calculated cross sections by the Nijmegen ESC08c model  based on the boson-exchange picture, the fss2 model including  QCM and the extended $\chi$EFT model, respectively.
  }
  \label{showAllSigmaMPdSdW3}
\end{figure*}

The differential cross sections were derived based on the equation (\ref{equ_cs}) where the number of scattering events were corrected by the CATCH efficiency depending on the $z$ vertex position.
In this case, the event count  $N_{scat}(i_{vtz}, \cos \theta)$ was obtained by subtracting the simulated background structure from the $\Delta E (\Sigma^{-}p)$ spectra in each $z$ vertex region.
The event count was then corrected by the corresponding efficiency of the $z$ vertex position.
We also calculated the differential cross section in a different way where the total event counts for all the vertex regions were corrected with the averaged efficiency for the $z$ vertex position (Fig. \ref{compSigmaMP_Eff_c4}).
In this method, the total event count was obtained from the counts of the fitted simulation spectrum for the $\Sigma^{-}p$ reaction in the $\Delta E (\Sigma^{-}p)$ spectrum.
We used two background structures obtained from the simultaneous fit of all $\Delta E$ and $\Delta p$ spectra (BG1) and single fit of  $\Delta E (\Sigma^{-}p)$ spectrum (BG2), as described in the previous subsection.

Fig. \ref{showAllSigmaMPdSdW3_1} shows the obtained differential cross sections for the $\Sigma^{-}p$ elastic scattering.
In this figure, the closed circles and open boxes were obtained from BG1 and BG2, respectively.
The red and blue colors show that the values were obtained with the $z$ vertex dependent efficiency and  $z$ vertex averaged efficiency, respectively.
The obtained values were almost consistent with each other.
However, there were some discrepancies around the edge region of the acceptance owing to the uncertainty of the background spectrum.
The mean values for each scattering angle were used as the differential cross section, and the root mean square was included in the systematic error in the background estimation.

Fig. \ref{showAllSigmaMPdSdW3} shows the measured differential cross sections with a past measurement (KEK-PS E289 data for 400 $< p$ (MeV/$c$) $<$ 700 \cite{Kondo:2000}) and theoretical calculations.
Here, the error bars and  black boxes show the statistical and systematic uncertainties, respectively.
The systematic uncertainty was estimated as the quadratic sum of the systematic error sources whose main components were the uncertainties of the background estimation in the $\Delta E(\Sigma^{-}p)$ spectra and the CATCH efficiency.
The differential cross section and its uncertainty are summarized in Table \ref{table_dsdw1}, \ref{table_dsdw2}, \ref{table_dsdw3} and  \ref{table_dsdw4} in Appendix \ref{appendix_2}, where  each source of the systematic error is also written.
Although the total error increases at the edge of the angular acceptance,  the data quality is drastically improved compared with the past experiments.
The data show a clear forward-peaking angular dependence for every momentum region, although  the statistical fluctuation increases in the momentum higher than 650 MeV/$c$.
Theoretical calculations for each momentum region are also overlaid.
The dotted (magenta), dot-dashed (blue) and solid (yellow) lines represent calculations by the Nijmegen ESC08c model based on the boson-exchange picture,  the fss2  model which includes  QCM, and the extended $\chi$EFT, respectively.
For every momentum region, the theoretical prediction by fss2 shows a rather good agreement with our data in the angular dependence and its absolute values, though there is a sizable difference at the lower momentum region.
The Nijmegen ESC08c model underestimates the differential cross sections at the forward angle.
For the $\chi$EFT with the error band owing to the cut-off value, the calculation with the cut-off value of 600 MeV/$c$ shown by the yellow solid line also underestimates the differential cross section especially in the lower momentum regions.
In the present model, the low-energy constants (LEC) representing the short-range part of the interaction were not well fixed owing to insufficient experimental data particularly for the high momentum ($>$ 300 MeV/$c$) region.
Our data will be an essential input to fix the LEC parameters in the extended $\chi$EFT model.

\section{Summary and Prospect} \label{sec_summary}

The study of the $YN$ interactions is important to expand our knowledge on the $NN$ interaction to the generalized $BB$ interactions within the SU(3) flavor symmetry.
The $YN$ interactions are also the basic information for the nuclear system with hyperons such as the hypernuclei and neutron stars.
Scattering experiments between a hyperon and  proton, which have been experimentally difficult up to the present due to the hyperon's short lifetime, are essential to test and improve the theoretical models of the $BB$ interactions.
To realize the high-statistics hyperon-proton scattering experiment, we performed the $\Sigma p$ scattering experiment at J-PARC with a high intensity $\pi$ beam to measure the differential cross sections with  better accuracy.
In this experiment, we measured the differential cross sections of the $\Sigma^{+}p, \Sigma^{-}p $, and $\Sigma^{-}p \rightarrow \Lambda n $ reactions to study the $\Sigma N$ interaction systematically.
In this paper, we present the results of the $\Sigma^{-}p$ elastic scattering.
Data for the momentum-tagged $\Sigma^{-}$ particles running in the LH$_{2}$ target were accumulated by detecting the $\pi^{-}p \to K^{+} \Sigma^{-}$ reaction with a high intensity $\pi^{-}$ beam of 20 M/spill.
In total, 1.62$\times$ $10^{7}$ $\Sigma^{-}$ particles were accumulated, and the momentum of the $\Sigma^{-}$ particles ranged from 470 to 850 MeV/$c$.
The CATCH system surrounding the LH$_{2}$ target was employed to detect a recoil proton and $\pi^{-}$ from the $\Sigma^{-}$ decay.
The $\Sigma^{-}p$ elastic scattering events were identified kinematically from the energy and angular information of these particles in the final state.
To derive the differential cross sections, the CATCH efficiency  was carefully estimated from the $pp$ scattering data with several different proton beam momenta.
We successfully measured the differential cross sections of the $\Sigma^{-}p$ elastic scattering for the momentum region from 470 to 850 MeV/$c$.
The statistical error of  the 10\% level was achieved  with a fine angular step of $d\cos \theta = 0.1$ by identifying  the largest ever statistics of approximately 4,500 $\Sigma^{-}p$ elastic scattering  events from 1.72 $\times$ $10^{7}$ $\Sigma^{-}$ particles.
The differential cross sections show a clear forward peaking structure, and the forward and backward ratios are large particularly in the higher momentum regions.
The experimental inputs of the two-body hyperon-proton scattering were quite limited up to now.
However the success of the $\Sigma^{-}p$ scattering experiment is a major breakthrough in providing accurate data to improve the $BB$ interaction models and  establish realistic $BB$ interactions.
Analysis to derive the differential cross sections of the $\Sigma^{-}p \to \Lambda n$ reaction and $\Sigma^{+}p$ elastic scattering is ongoing.
Because all channels are related to each other within the framework of flavor SU(3) symmetry, these data also impose strong constraints on the theories of two-body $BB$ interaction.
By combining all the experimental information, a better understanding of the $BB$ interactions will be achieved in near future.

\begin{acknowledgments}

We would like to thank the staff of the J-PARC accelerator and Hadron Experimental Facility for their support to provide the beam during the beam time.
We also thank the staff of CYRIC and ELPH at Tohoku University for their support to provide beams for the test experiment for our detectors.
We would like to express gratitude to  Y. Fujiwara for the theoretical support from the initial stages of the experimental design and  thank T. A. Rijken and J. Haidenbauer for their theoretical calculations.
We also thank KEKCC and  SINET4.
This work was supported by  JSPS KAKENHI Grant Number 23684011, 15H00838, 15H05442, 15H02079 and 18H03693.
This work was also supported by  Grants-in-Aid Number 24105003 and 18H05403 for Scientific Research from the Ministry of Education, Culture, Science and Technology (MEXT) Japan.

\end{acknowledgments}

\appendix
\section{Angular dependence of the $\Delta E (\Sigma^{-}p)$ distribution}\label{appendix_1}
\begin{figure*}[]
  \centerline{\includegraphics[width=0.6\textwidth]{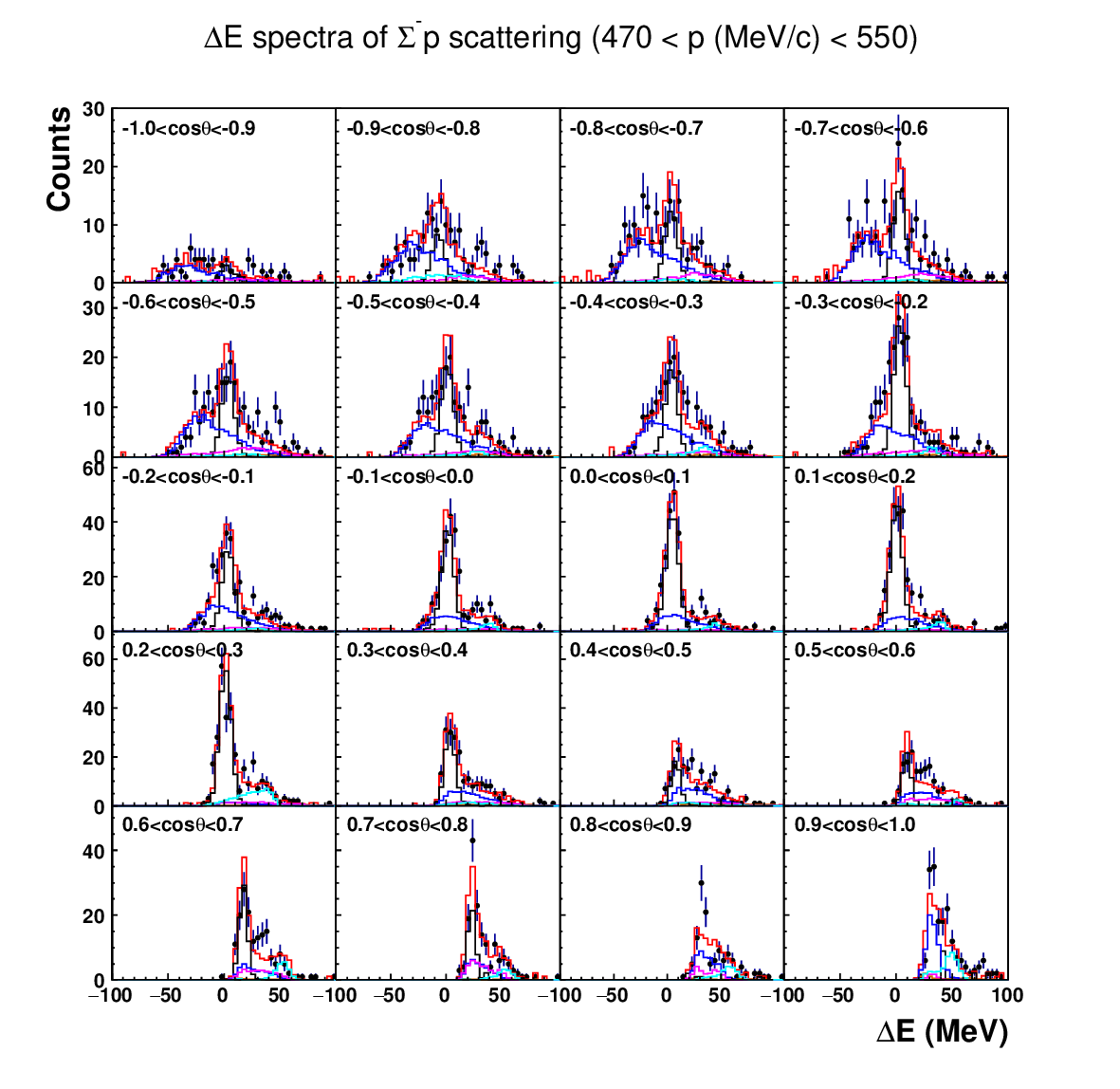}}
  \centerline{\includegraphics[width=0.6\textwidth]{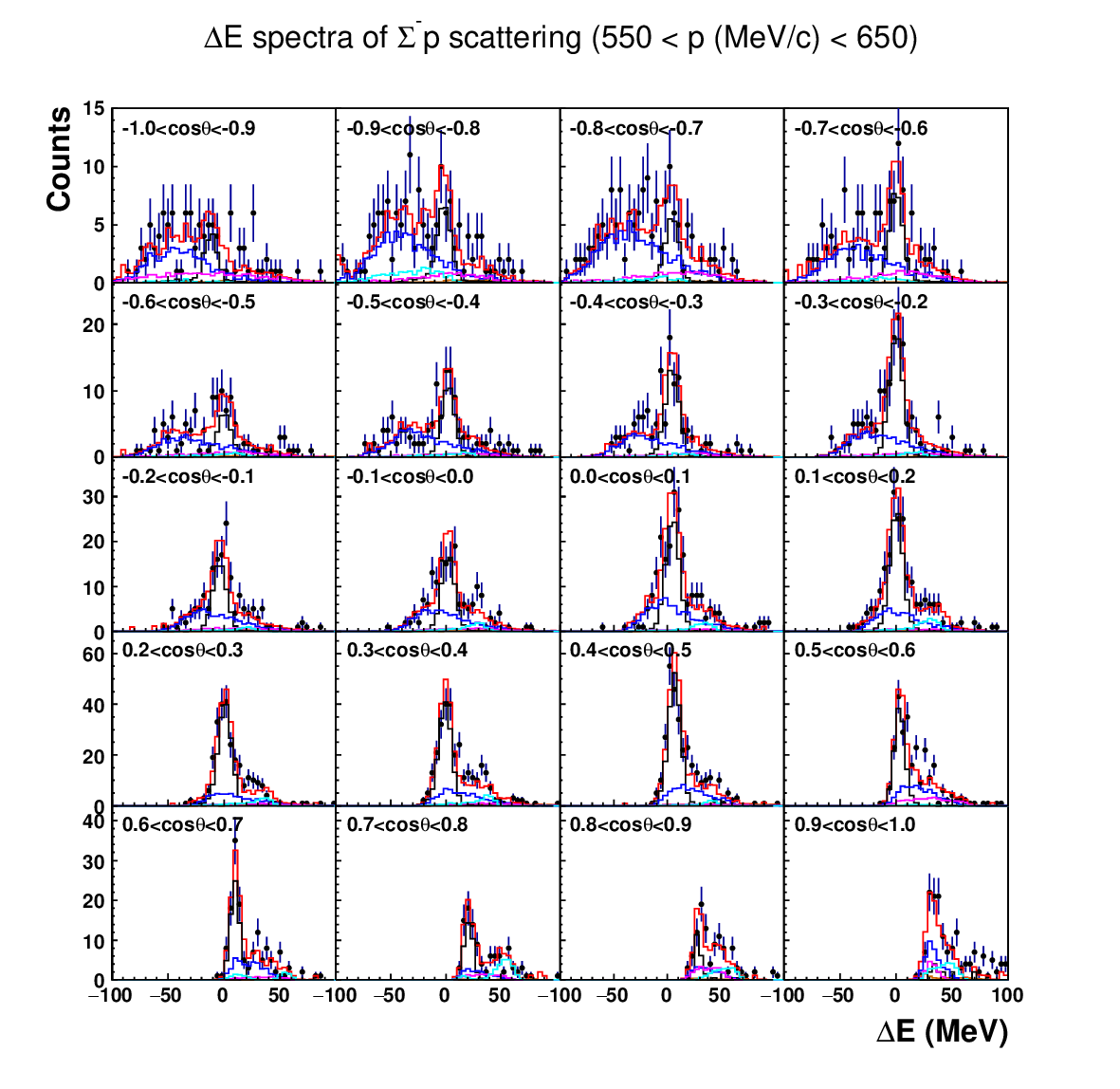}}
  \caption{ $\Delta E (\Sigma^{-}p)$ spectra for each scattering angle for 0.45 $<$ $p$ (GeV/$c$) $<$ 0.55 (top) and 0.55 $<$ $p$ (GeV/$c$) $<$ 0.65 (bottom).
  The data points and histograms are the same as those in Fig. \ref{showSigmaMPScatWithFit4_wBG_result_0_c1}.
  }
  \label{showSigmaMPScatWithFit4_result_0_c5}
\end{figure*}

\begin{figure*}[]
  \centerline{\includegraphics[width=0.6\textwidth]{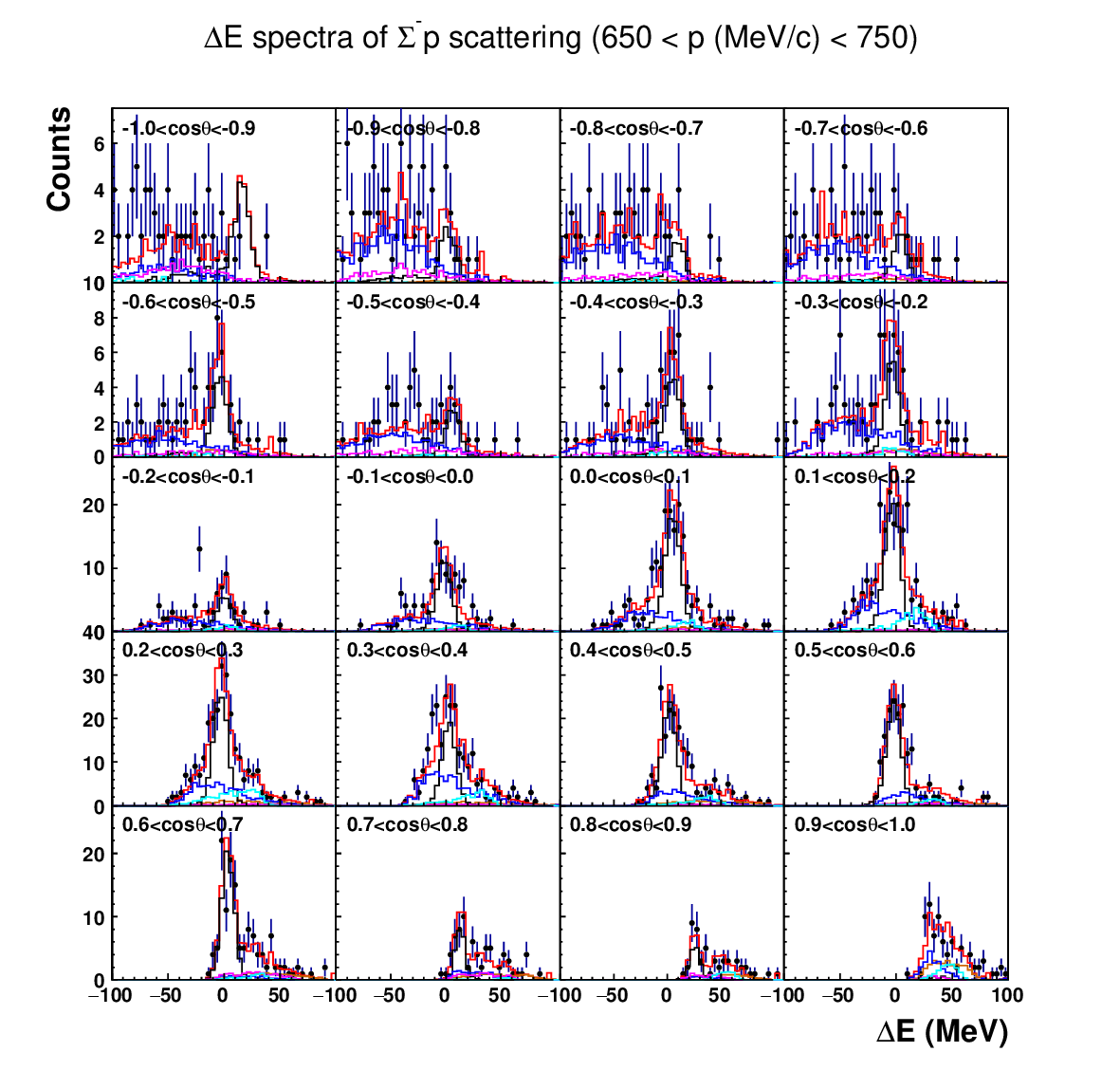}}
  \centerline{\includegraphics[width=0.6\textwidth]{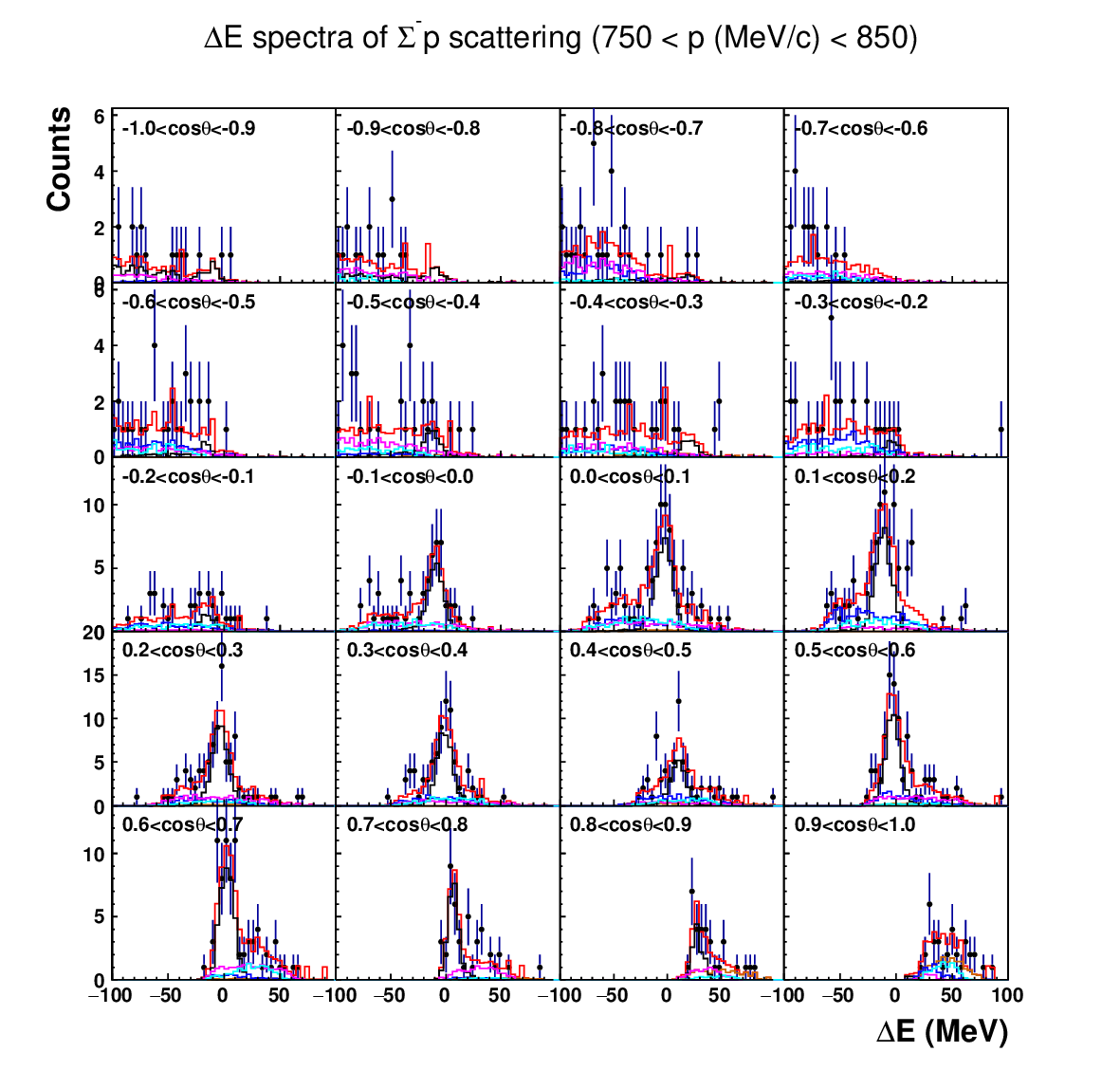}}
  \caption{ $\Delta E (\Sigma^{-}p)$ spectra for each scattering angle for 0.65 $<$ $p$ (GeV/$c$) $<$ 0.75 (top) and 0.75 $<$ $p$ (GeV/$c$) $<$ 0.85 (bottom).
  The data points and histograms are the same as those in Fig. \ref{showSigmaMPScatWithFit4_wBG_result_0_c1}.
  }
  \label{showSigmaMPScatWithFit4_result_2_c5}
\end{figure*}

 The $\Delta E(\Sigma^{-}p)$ spectra are shown in Fig.  \ref{showSigmaMPScatWithFit4_result_0_c5} and \ref{showSigmaMPScatWithFit4_result_2_c5} for each scattering angle for the four momentum regions to show the statistical significance of the $\Sigma^{-}p$ scattering events.
 The fit results with the sum of the simulated spectra of the $\Sigma^{-}p$ scattering and background contributions are also shown.
The histogram color for each component is the same with the ones in Fig. \ref{showSigmaMPScatWithFit4_wBG_result_0_c1}.
In every angular region, the $\Delta E(\Sigma^{-}p)$ spectrum can be reproduced by the sum of the simulated spectra.
 Because the $\Sigma^{-}p \to \Sigma^{0}n$ reaction cannot be removed from the kinematical analysis, the background contribution from this reaction remains in the $\Delta E (\Sigma^{-}p)$ spectrum.

\section{Table of the differential cross section } \label{appendix_2}

We summarize the derived differential cross section values and its uncertainties in Tables \ref{table_dsdw1}, \ref{table_dsdw2} , \ref{table_dsdw3}  and \ref{table_dsdw4}  for the four momentum regions.
The statistical and systematic errors are listed in the columns of the stat. and syst. (total), respectively.
The systematic error is estimated as a quadratic sum of the error sources from the background estimation (syst. (BG)),  CATCH efficiency (syst. (eff)) and $\Sigma^{-}$ total flight length (syst. ($L$)).

\begin{table*}
\begin{center}
\caption{Table of the differential cross section of the $\Sigma^{-}p$ elastic scattering for 470 $<$ $p_{\Sigma^{-}}$(MeV/$c$) $<$ 550 MeV.
The statistical and systematic errors are listed in the stat. and syst. (total) columns, respectively.
The systematic error is estimated as a quadratic sum of the error sources from the background estimation (syst. (BG)),  CATCH efficiency (syst. (eff)) and $\Sigma^{-}$ total flight length (syst. ($L$)).
}
\label{table_dsdw1}
\begin{tabular}{cccc|ccc}
\\
\hline\hline
$\cos \theta$  &  $\frac{d\sigma}{d\Omega}$ &  stat. &  syst. (total) &  syst. (BG) &  syst. (eff) & syst. ($L$)\\
                     &  (mb/sr) &  (mb/sr) &  (mb/sr) &  (mb/sr)&  (mb/sr) &  (mb/sr) \\
\hline
$-$0.95$\pm$0.05 &  0.54 &  $\pm$0.41 &  $^{+0.40}_{-0.40}$  &  $\pm$0.403 &   $^{+0.004}_{-0.0}$  &  $\pm$0.004 \\
$-$0.85$\pm$0.05 &  1.01 &  $\pm$0.36 &  $^{+0.16}_{-0.16}$  &  $\pm$0.159 &   $^{+0.003}_{-0.000}$  &  $\pm$0.008 \\
$-$0.75$\pm$0.05 &  0.53 &  $\pm$0.24 &  $^{+0.25}_{-0.25}$  &  $\pm$0.252 &   $^{+0.001}_{-0.000}$  &  $\pm$0.004 \\
$-$0.65$\pm$0.05 &  0.71 &  $\pm$0.17 &  $^{+0.13}_{-0.13}$  &  $\pm$0.134 &   $^{+0.0}_{-0.002}$  &  $\pm$0.005 \\
$-$0.55$\pm$0.05 &  0.62 &  $\pm$0.15 &  $^{+0.10}_{-0.10}$  &  $\pm$0.103 &   $^{+0.004}_{-0.000}$  &  $\pm$0.004 \\
$-$0.45$\pm$0.05 &  0.61 &  $\pm$0.15 &  $^{+0.13}_{-0.13}$  &  $\pm$0.127 &   $^{+0.006}_{-0.0}$  &  $\pm$0.004 \\
$-$0.35$\pm$0.05 &  0.64 &  $\pm$0.14 &  $^{+0.06}_{-0.06}$  &  $\pm$0.058 &   $^{+0.002}_{-0.003}$  &  $\pm$0.005 \\
$-$0.25$\pm$0.05 &  1.01 &  $\pm$0.16 &  $^{+0.08}_{-0.08}$  &  $\pm$0.084 &   $^{+0.0}_{-0.005}$  &  $\pm$0.008 \\
$-$0.15$\pm$0.05 &  1.13 &  $\pm$0.19 &  $^{+0.12}_{-0.12}$  &  $\pm$0.122 &   $^{+0.008}_{-0.020}$  &  $\pm$0.009 \\
$-$0.05$\pm$0.05 &  1.48 &  $\pm$0.20 &  $^{+0.08}_{-0.11}$  &  $\pm$0.077 &   $^{+0.020}_{-0.082}$  &  $\pm$0.011 \\
0.05$\pm$0.05 &  1.87 &  $\pm$0.23 &  $^{+0.08}_{-0.17}$  &  $\pm$0.053 &   $^{+0.051}_{-0.157}$  &  $\pm$0.014 \\
0.15$\pm$0.05 &  2.31 &  $\pm$0.28 &  $^{+0.16}_{-0.27}$  &  $\pm$0.127 &   $^{+0.100}_{-0.239}$  &  $\pm$0.018 \\
0.25$\pm$0.05 &  3.34 &  $\pm$0.31 &  $^{+0.31}_{-0.40}$  &  $\pm$0.278 &   $^{+0.126}_{-0.286}$  &  $\pm$0.026 \\
0.35$\pm$0.05 &  2.44 &  $\pm$0.44 &  $^{+0.21}_{-0.21}$  &  $\pm$0.188 &   $^{+0.084}_{-0.090}$  &  $\pm$0.019 \\
\hline\hline
\end{tabular}
\end{center}
\end{table*}

\begin{table*}
\begin{center}
\caption{Table of the differential cross section of the $\Sigma^{-}p$ elastic scattering for 550 $<$ $p_{\Sigma^{-}}$(MeV/$c$) $<$ 650 MeV.
}
\label{table_dsdw2}
\begin{tabular}{cccc|ccc}
\\
\hline\hline
$\cos \theta$  &  $\frac{d\sigma}{d\Omega}$ &  stat.  &  syst. (total) &  syst. (BG) &  syst. (eff)  & syst. ($L$)\\
                     &  (mb/sr) &  (mb/sr) &  (mb/sr) &  (mb/sr)&  (mb/sr) &  (mb/sr) \\
\hline
$-$0.95$\pm$0.05 &  0.84 &  $\pm$0.66 &  $^{+0.41}_{-0.41}$  &  $\pm$0.411 &   $^{+0.012}_{-0.0}$  &  $\pm$0.003 \\
$-$0.85$\pm$0.05 &  0.78 &  $\pm$0.40 &  $^{+0.34}_{-0.34}$  &  $\pm$0.339 &   $^{+0.0}_{-0.007}$  &  $\pm$0.003 \\
$-$0.75$\pm$0.05 &  0.53 &  $\pm$0.31 &  $^{+0.17}_{-0.17}$  &  $\pm$0.166 &   $^{+0.0}_{-0.001}$  &  $\pm$0.002 \\
$-$0.65$\pm$0.05 &  0.72 &  $\pm$0.24 &  $^{+0.12}_{-0.12}$  &  $\pm$0.118 &   $^{+0.0}_{-0.004}$  &  $\pm$0.002 \\
$-$0.55$\pm$0.05 &  0.67 &  $\pm$0.19 &  $^{+0.02}_{-0.02}$  &  $\pm$0.020 &   $^{+0.004}_{-0.003}$  &  $\pm$0.002 \\
$-$0.45$\pm$0.05 &  0.92 &  $\pm$0.19 &  $^{+0.03}_{-0.03}$  &  $\pm$0.030 &   $^{+0.007}_{-0.005}$  &  $\pm$0.003 \\
$-$0.35$\pm$0.05 &  1.12 &  $\pm$0.21 &  $^{+0.06}_{-0.06}$  &  $\pm$0.059 &   $^{+0.016}_{-0.004}$  &  $\pm$0.004 \\
$-$0.25$\pm$0.05 &  1.35 &  $\pm$0.22 &  $^{+0.09}_{-0.09}$  &  $\pm$0.088 &   $^{+0.005}_{-0.005}$  &  $\pm$0.005 \\
$-$0.15$\pm$0.05 &  1.15 &  $\pm$0.21 &  $^{+0.02}_{-0.02}$  &  $\pm$0.023 &   $^{+0.0}_{-0.008}$  &  $\pm$0.004 \\
$-$0.05$\pm$0.05 &  1.01 &  $\pm$0.24 &  $^{+0.13}_{-0.13}$  &  $\pm$0.127 &   $^{+0.017}_{-0.0}$  &  $\pm$0.004 \\
0.05$\pm$0.05 &  1.69 &  $\pm$0.25 &  $^{+0.03}_{-0.03}$  &  $\pm$0.026 &   $^{+0.0}_{-0.015}$  &  $\pm$0.006 \\
0.15$\pm$0.05 &  1.72 &  $\pm$0.33 &  $^{+0.25}_{-0.25}$  &  $\pm$0.248 &   $^{+0.007}_{-0.032}$  &  $\pm$0.006 \\
0.25$\pm$0.05 &  2.52 &  $\pm$0.36 &  $^{+0.28}_{-0.30}$  &  $\pm$0.274 &   $^{+0.059}_{-0.121}$  &  $\pm$0.010 \\
0.35$\pm$0.05 &  3.26 &  $\pm$0.45 &  $^{+0.49}_{-0.56}$  &  $\pm$0.486 &   $^{+0.072}_{-0.288}$  &  $\pm$0.013 \\
0.45$\pm$0.05 &  5.35 &  $\pm$0.76 &  $^{+0.86}_{-0.94}$  &  $\pm$0.838 &   $^{+0.178}_{-0.422}$  &  $\pm$0.021 \\
0.55$\pm$0.05 &  5.19 &  $\pm$0.71 &  $^{+0.83}_{-0.61}$  &  $\pm$0.601 &   $^{+0.571}_{-0.103}$  &  $\pm$0.020 \\
\hline\hline
\end{tabular}
\end{center}
\end{table*}

\begin{table*}
\begin{center}
\caption{Table of the differential cross section of the $\Sigma^{-}p$ elastic scattering for 650 $<$ $p_{\Sigma^{-}}$(MeV/$c$) $<$ 750 MeV.
}
\label{table_dsdw3}
\begin{tabular}{cccc|ccc}
\\
\hline\hline
$\cos \theta$  &  $\frac{d\sigma}{d\Omega}$ &  stat. &  syst. (total) &  syst. (BG) &  syst. (eff) &  syst. ($L$) \\
                     &  (mb/sr) &  (mb/sr) &  (mb/sr) &  (mb/sr)&  (mb/sr) &  (mb/sr) \\
\hline
$-$0.75$\pm$0.05 &  0.45 &  $\pm$0.40 &  $^{+0.15}_{-0.15}$  &  $\pm$0.154 &   $^{+0.009}_{0.0}$  &  $\pm$0.001 \\
$-$0.65$\pm$0.05 &  0.44 &  $\pm$0.37 &  $^{+0.15}_{-0.15}$  &  $\pm$0.153 &   $^{+0.005}_{0.0}$  &  $\pm$0.001 \\
$-$0.55$\pm$0.05 &  1.14 &  $\pm$0.46 &  $^{+0.13}_{-0.12}$  &  $\pm$0.124 &   $^{+0.029}_{0.0}$  &  $\pm$0.004 \\
$-$0.45$\pm$0.05 &  0.45 &  $\pm$0.24 &  $^{+0.07}_{-0.07}$  &  $\pm$0.074 &   $^{+0.010}_{0.0}$  &  $\pm$0.001 \\
$-$0.35$\pm$0.05 &  0.99 &  $\pm$0.30 &  $^{+0.10}_{-0.10}$  &  $\pm$0.100 &   $^{+0.003}_{-0.008}$  &  $\pm$0.003 \\
$-$0.25$\pm$0.05 &  0.76 &  $\pm$0.31 &  $^{+0.10}_{-0.10}$  &  $\pm$0.100 &   $^{+0.004}_{-0.015}$  &  $\pm$0.003 \\
$-$0.15$\pm$0.05 &  0.67 &  $\pm$0.30 &  $^{+0.11}_{-0.11}$  &  $\pm$0.107 &   $^{+0.003}_{-0.006}$  &  $\pm$0.002 \\
$-$0.05$\pm$0.05 &  1.39 &  $\pm$0.32 &  $^{+0.18}_{-0.18}$  &  $\pm$0.178 &   $^{+0.019}_{-0.0}$  &  $\pm$0.005 \\
0.05$\pm$0.05 &  2.42 &  $\pm$0.46 &  $^{+0.12}_{-0.12}$  &  $\pm$0.114 &   $^{+0.023}_{-0.0}$  &  $\pm$0.009 \\
0.15$\pm$0.05 &  2.12 &  $\pm$0.43 &  $^{+0.18}_{-0.18}$  &  $\pm$0.176 &   $^{+0.0}_{-0.031}$  &  $\pm$0.008 \\
0.25$\pm$0.05 &  3.07 &  $\pm$0.43 &  $^{+0.16}_{-0.16}$  &  $\pm$0.155 &   $^{+0.010}_{-0.0}$  &  $\pm$0.012 \\
0.35$\pm$0.05 &  2.21 &  $\pm$0.45 &  $^{+0.17}_{-0.17}$  &  $\pm$0.165 &   $^{+0.029}_{-0.001}$  &  $\pm$0.008 \\
0.45$\pm$0.05 &  2.73 &  $\pm$0.61 &  $^{+0.29}_{-0.34}$  &  $\pm$0.293 &   $^{+0.006}_{-0.174}$  &  $\pm$0.010 \\
0.55$\pm$0.05 &  4.44 &  $\pm$0.61 &  $^{+0.37}_{-0.56}$  &  $\pm$0.348 &   $^{+0.117}_{-0.438}$  &  $\pm$0.017 \\
0.65$\pm$0.05 &  5.43 &  $\pm$0.79 &  $^{+0.50}_{-0.38}$  &  $\pm$0.317 &   $^{+0.394}_{-0.206}$  &  $\pm$0.021 \\
\hline\hline
\end{tabular}
\end{center}
\end{table*}

\begin{table*}
\begin{center}
\caption{Table of the differential cross section of the $\Sigma^{-}p$ elastic scattering for 750 $<$ $p_{\Sigma^{-}}$(MeV/$c$) $<$ 850 MeV.
}
\label{table_dsdw4}
\begin{tabular}{cccc|ccc}
\\
\hline\hline
$\cos \theta$  &  $\frac{d\sigma}{d\Omega}$ &  stat.  &  syst. (total) &  syst. (BG) &  syst. (eff) &  syst. ($L$) \\
                     &  (mb/sr) &  (mb/sr) &  (mb/sr) &  (mb/sr) &  (mb/sr) &  (mb/sr) \\
\hline
$-$0.25$\pm$0.05 &  0.54 &  $\pm$0.52 &  $^{+0.13}_{-0.13}$  &  $\pm$0.127 &   $^{+0.0}_{-0.019}$  &  $\pm$0.002 \\
$-$0.15$\pm$0.05 &  0.83 &  $\pm$0.43 &  $^{+0.17}_{-0.17}$  &  $\pm$0.173 &   $^{+0.0}_{-0.017}$  &  $\pm$0.003 \\
$-$0.05$\pm$0.05 &  2.25 &  $\pm$0.62 &  $^{+0.17}_{-0.17}$  &  $\pm$0.172 &   $^{+0.009}_{-0.0}$  &  $\pm$0.008 \\
0.05$\pm$0.05 &  2.81 &  $\pm$0.64 &  $^{+0.07}_{-0.07}$  &  $\pm$0.047 &   $^{+0.054}_{-0.050}$  &  $\pm$0.011 \\
0.15$\pm$0.05 &  3.20 &  $\pm$0.66 &  $^{+0.17}_{-0.17}$  &  $\pm$0.170 &   $^{+0.008}_{-0.043}$  &  $\pm$0.012 \\
0.25$\pm$0.05 &  3.02 &  $\pm$0.67 &  $^{+0.12}_{-0.11}$  &  $\pm$0.113 &   $^{+0.044}_{-0.0}$  &  $\pm$0.012 \\
0.35$\pm$0.05 &  2.57 &  $\pm$0.58 &  $^{+0.16}_{-0.15}$  &  $\pm$0.154 &   $^{+0.037}_{-0.0}$  &  $\pm$0.010 \\
0.45$\pm$0.05 &  1.82 &  $\pm$0.61 &  $^{+0.15}_{-0.16}$  &  $\pm$0.150 &   $^{+0.0}_{-0.045}$  &  $\pm$0.007 \\
0.55$\pm$0.05 &  3.38 &  $\pm$0.82 &  $^{+0.15}_{-0.21}$  &  $\pm$0.145 &   $^{+0.0}_{-0.156}$  &  $\pm$0.013 \\
0.65$\pm$0.05 &  4.65 &  $\pm$0.92 &  $^{+0.38}_{-0.51}$  &  $\pm$0.323 &   $^{+0.201}_{-0.387}$  &  $\pm$0.018 \\
0.75$\pm$0.05 &  5.59 &  $\pm$1.60 &  $^{+1.24}_{-0.50}$  &  $\pm$0.502 &   $^{+1.133}_{-0.0}$  &  $\pm$0.022 \\
\hline\hline
\end{tabular}
\end{center}
\end{table*}

\bibliography{bibliography}

\end{document}